\documentstyle[preprint,aps,eqsecnum]{revtex}

\newcommand{\be}{\begin{eqnarray}}
\newcommand{\ee}{\end{eqnarray}}
\newcommand{\BE} {\begin{equation}}
\newcommand{\EE} {\end{equation}}

\begin{document}
\draft

\title{
\begin{flushright}
{\normalsize IP-ASTP-19-95}
\end{flushright}
\Large\bf Quark Confinement in Light-Front QCD \\
		and\\
	A Weak-Coupling Treatment to Heavy Hadrons}

\author{{\bf Wei-Min Zhang}\thanks{
		E-Mail address: wzhang@phys.sinica.edu.tw} \\
	Institute of Physics, Academia Sinica, Taipei 11529, Taiwan}
\date{October 25, 1995}

\maketitle

\begin{abstract}
In this paper, we develop a weak-coupling treatment of
nonperturbative QCD to heavy hadrons on the light-front.
First, we present a derivation of quark confining interaction
in light-front QCD for heavy quark systems, based on the
recently developed light-front similarity renormalization group
approach and the light-front heavy quark effective theory.
The resulting effective light-front QCD Hamiltonian $H_\lambda$
at a low-energy cutoff $\lambda$ manifests the coexistence of a
confining potential and a Coulomb potential. A clear light-front
picture of quark confinement emerges.  Using this low energy
QCD Hamiltonian $H_\lambda$, we study heavy hadron bound state
equations in the framework of a recently proposed possible
weak-coupling treatment of non-perturbative QCD.
Light-front heavy hadron bound states with definite spin
and parity are constructed and the general structure of the
corresponding wavefunctions is explored. A Gaussian-type
wavefunction ansatz is used to solve the light-front
quarkonium bound state equation.  We find that the effective
coupling constant determined from the quarkonium bound state equation
can be arbitrarily small so that the weak-coupling treatment
to heavy hadron bound states in light-front QCD is explicitly
achieved. Finally, the scale dependence of the effective
coupling constant is analytically calculated and the similarity
renormalization group $\beta$ function is determined, from
which the running coupling constant in small momentum transfer
is given qualitatively by $\overline{\alpha}(Q^2) \sim
{\Lambda_{QCD}^2 \over Q^2}$.  Such a running coupling constant is
the basic assumption in the successful Richardson $Q\overline{Q}$
potential that ensures the existence of a linear confining
potential at large distance, but now can be obtained
from light-front QCD.
\end{abstract}

\vspace{0.5in}


\newpage


\section{Introduction}

There are two fundamental problems in QCD for hadronic physics,
the quark confinement and the spontaneous breaking of chiral
symmetry. These two problems are the basis for solving the
low-energy hadronic bound states from QCD but none of them
has been completely understood.  Recently, Wilson and
his collaborators proposed an approach to determine
the low-energy bound states in nonperturbative QCD as a
weak-coupling problem \cite{Wilson94}.
The key to eliminating necessarily nonperturbative effects
is to construct a low-energy QCD Hamiltonian in which quarks
and gluons have nonzero constituent masses rather than the
zero masses of the current picture.  The use of constituent
masses cuts off the growth of the running coupling constant
and makes it conceivable that the running coupling never
leaves the perturbative domain. The weak-coupling approach
potentially reconciles the simplicity of the constituent quark
model with the complexities of QCD. The penalty for achieving
this weak-coupling picture is the necessity of formulating
the problem in light-front coordinates and of dealing with
the complexities of renormalization. To handle the complexities
of light-front renormalization, a new renormalization approach,
so-called the similarity renormalization group scheme, has also
recently been developed \cite{Wilson94,Glazek94}.

Based on the idea of the light-front similarity renormalization
group scheme and the concept of coupling coherence
\cite{Perry93}, Perry has shown that upon a
calculation to the second order, there exists
a logarithmic confining potential in the resulting
light-front QCD effective Hamiltonian \cite{Perry94}.
This is a crucial finding for a practical realization of
the weak-coupling treatment to nonperturbative QCD.
However, the general strategy of solving hadrons through
the weak-coupling treatment scheme is far to be completed.
In this paper, we shall use the similarity renormalization
group approach to derive a low energy heavy quark QCD Hamiltonian,
based on the light-front heavy quark effective theory we developed
recently \cite{Zhang95,Cheung95}. The resulting Hamiltonian
exhibits explicitly the confining and Coulomb interactions.
We thereby adopt the idea of the weak-coupling approach
to study the strongly interacting heavy hadronic structure
on the light-front.  From this investigation we explicitly
provide a realization of the weak-coupling treatment to
nonperturbative QCD.

The reason of choosing heavy hadron systems as a
starting example is clear: For light quark
systems, both  quark confinement and  spontaneous
chiral symmetry breaking play an essential role to the quark
dynamics in hadrons. However, the success of the chiral symmetry
description of the low energy hadron physics naturally
indicates a chiral symmetry breaking scale ($\Lambda_{\chi SB}
\sim 1~GeV$) which is relatively larger than the confinement
scale ($\Lambda_{QCD} \sim 200~MeV$).  In other words, for
light quark systems, it seems to be necessary to understand
the underlying mechanism of chiral symmetry breaking
before we can further explore the mechanism of quark
confinement. Of course, for the best description,
both problems should be solved simultaneously in the same picture,
but at the moment, this will certainly complicate the
study on confinement. It would be nice if we could
separately deal with these two most difficult
but fundamental problems in QCD.
For heavy quark systems, chiral symmetry is explicitly
broken so that confinement is the sole nontrivial
feature influencing heavy quark dynamics.

Confinement interaction may not be very sensitive but it is
important in describing the QCD dynamics of quarkonium
spectroscopy and their decay processes. And it should play
a more important role in heavy-light quark systems, such
as $B$ and $D$ mesons. One may argue that the mass scales for
heavy and light quark systems are different. The heavy quark energy
cannot run down to the usual hadronic scale of light quark systems due
to heavy quark mass.  Meanwhile, confining interactions must be
energy scale dependent.  Apparently, confinement for heavy quark
systems could be very different from light quark systems.
However, despite heavy and light quarks, confinement arises only
from low energy gluon interactions.  In other words, the
confinement mechanism should be the same for both heavy and
light quark systems.   We thus choose heavy hadron systems
without any loss of generality.  In order to avoid the possible
confusion about the different mass scales and to correctly extract
the confining interactions in low energy heavy quark dynamics,
it is convenient to work with
heavy quark effective theory (HQET).  The HQET
is a theory of QCD in $1/m_Q$ expansion \cite{Georgi}, where
$m_Q$ is the heavy quark mass. In HQET,
the low-energy dynamics is determined through
the interacting gluons and heavy quarks by exchanging a small
residual momentum of heavy quarks, which
is of order $\Lambda_{QCD}$. As a result, within HQET we can
indeed explore the low energy QCD dynamics for heavy quark systems
in the same scale as that for light quark systems. Meanwhile, the
extension of the study to light quark systems becomes obviously
straightforward, although undoubtedly the corresponding result
must be very complicated due to the spin dependence of the low energy
interacting Hamiltonian. The spin dependent interactions on the
light-front are essentially related to the chiral symmetry breaking.
These spin dependent interactions in HQET are suppressed
in the leading order approximation because they are
$1/m_Q$ corrections and can be treated perturbatively
with respect to the heavy hadron states.  This is why
for heavy quark systems the chiral symmetry
breaking can be treated separately from the confinement.

In fact, the model-based theoretical investigations on heavy
quarkonia lasted for one and half decades is recently replacing by
first-principles exploration on QCD.  The lattice QCD simulation
may give an acceptable description for heavy
quarkonium spectroscopy with manageable control over all
the systematic errors \cite{Lepage91}. The development of
nonrelativistic QCD provides a general factorization formula
to quarkonium annihilation and production processes so that
a rigorous QCD analysis may become possible \cite{Lepage95}.
Meanwhile, in the past five years considerable progress has been made
for heavy hadrons with one heavy quark, due mainly to the discovery
of the so-called  heavy quark symmetry (HQS) \cite{Isgur90}
 and the development
of the heavy quark effective theory (HQET)\cite{Georgi} from QCD.
The HQS and HQET have in certain contents put the description
of heavy hadron physics on a QCD-related and model-independent
basis. Yet, a truly first-principles QCD understanding
of heavy hadrons is still lacking since no good nonperturbative
QCD approach is available for a direct computation of heavy hadron
wavefunctions. On the other hand, in the last decade, the
investigations of the light-front field theory on nonperturbative
bound state problems have made some progress \cite{Brodsky93} but
no real hadronic problem has been solved from which. Starting
with heavy hadrons may provide a possible explicit solution
of hadronic bound states in light-front QCD.

Simply speaking, the approach to achieve the QCD description of
hadronic bound states that we shall study in this paper can be
summarized as follows: Applying the similarity renormalization
group approach to light-front
QCD, we can obtain an low energy QCD Hamiltonian
which is an expansion in terms of the QCD coupling constant.
Then we attempt to solve from this low energy QCD Hamiltonian
 the strong interacting bound states as a weak-coupling
problem.  The weak-coupling treatment
contains the following steps: (i) Compute from the similarity
renormalization group scheme the low energy
Hamiltonian $H_\lambda$ at the low-energy cutoff $\lambda$ up
to the second order in coupling constant.
Then separate the Hamiltonian into a
nonperturbative part, $H_{\lambda 0}$ which contains
not only the free Hamiltonians of quarks
and gluons but also the dominant two-body interactions, and
the remaining part plus the higher order contributions generated
in the similarity renormalization as a perturbative term,
$H_{\lambda I}$. (ii) Introduce a constituent picture
which is an important step in the realization of the
weak-coupling treatment of nonperturbative QCD.
The constituent quarks and gluons have masses of
a few hundreds MeV, and these masses are functions of the cutoff
$\lambda$ that must vanish when the effective theory
goes back to the full QCD theory.
(iii) Solve hadronic bound states with $H_{\lambda 0}$
nonperturbatively in the constituent picture and
determine the cutoff dependence of the constituent
masses and the coupling constant.
The coupling constant $g$ now becomes an effective one,
$g_\lambda$. In the nonperturbative study
of $H_{\lambda 0}$, if we could show that with a suitable
choice of the low energy cutoff $\lambda$, the effective
coupling constant $g_\lambda$ is arbitrarily small, then
a weak-coupling treatment could be applied to the low
energy QCD $H_\lambda$ such that the corrections from
$H_{\lambda I}$ can really be computed perturbatively.
(iv) There should be a limit
$g_\lambda \rightarrow g_s$, where $g_s$ is the fixed
physical coupling constant measured at the hadronic
mass scale, such that all the constituent quarks and gluons become
current ones again. Then the effective low energy theory returns
back to the full QCD theory.
If everything listed above works well, we arrive at a
true weak-coupling QCD theory of the strong interaction for
hadrons.

In the following, I begin with a general bare interacting
Hamiltonian with a high energy cutoff $\Lambda$ that removes
the usual ultraviolet (UV) divergences.
Then using the light-front similarity
renormalization group, we construct a Hamiltonian
under the low energy cutoff $\lambda$. The low energy cutoff is
introduced via a smearing function in the similarity
renormalization group that effectively integrates over all
the modes above the cutoff $\lambda$.  The choice of the smearing
function in this paper is much simpler in comparison to the original
setup \cite{Wilson94}. Applying this general formula to the
light-front heavy quark effective theory (light-front HQET)
we developed recently
\cite{Zhang95,Cheung95}, a low energy confining QCD
Hamiltonian can be explicitly obtained for heavy hadron systems.
Consequently, a clear light-front picture of quark confinement
emerges.  Furthermore, based on the idea of weak-coupling
approach, we use this confining Hamiltonian to
study the light-front heavy hadron bound states nonperturbatively,
and we can then provide an explicit description of the weak-coupling
treatment in the light-front HQET for heavy hadrons.

The present paper is organized as follows.  In Section 2, we
present a general procedure of constructing a low energy
light-front QCD Hamiltonian in the similarity renormalization
group scheme and discuss the possible existence of confining
interaction in such a derivation.  In Section 3, applying
the general procedure to the light-front HQET of QCD,
we further derive the heavy quark confining Hamiltonian
at the low-energy scale. In Section 4, the light-front
heavy hadronic bound state equations in the constituent picture
are developed within the scheme of the weak-coupling treatment. A
light-front picture of quark confinement for heavy hadrons is
illustrated in Section 5.  In Section 6, the heavy hadron bound
state equation is solved for quarkonia with a Gaussian-type
light-front wavefunction ansatz, from which the scale dependence
of effective coupling constant is determined as a solution
of the similarity renormalization group equation on the
quarkonium binding energy in Section 7.  The low energy running
coupling constant is also qualitatively obtained. In Section 8,
a connection between the low energy effective theory to the
full QCD theory is explored and the consistency is provided.
Finally, a summary is presented in Section 9.

\section{Low energy QCD Hamiltonian in Similarity
Renormalization Group Scheme}

We begin with the general formulation of the similarity
renormalization group approach to construct a low energy
QCD Hamiltonian. In general, for a given bare Hamiltonian,
$H^B = H_0^B + H_I^B$, where $H^B_0$ is a bare free Hamiltonian
and $E_i$ is assumed to be its eigenvalue, the similarity
renormalization group approach leads to the following
Hamiltonian at the low energy cutoff $\lambda$ (for a detailed
derivation, see Ref. \cite{Wilson94}):
\begin{eqnarray}
	H_{\lambda} &=& \Bigg( H^B_{0\lambda} + H^B_{I\lambda} \Bigg) +
		\Bigg( {\underbrace{[H^B_{I\lambda^\prime},
		H^B_{I\lambda^\prime T}]}}_R \Bigg) \nonumber \\
	 && + \Bigg( {\underbrace{[{\underbrace{[H^B_{I\lambda^{
		\prime\prime}}, H^B_{I\lambda^{\prime\prime}
		T}]}}_{R^\prime} , H^B_{I\lambda^\prime T}]}}_R
		+{\underbrace{[H^B_{I\lambda^\prime},
		{\underbrace{[H^B_{I\lambda^{\prime\prime}},
		H^B_{I\lambda^{\prime\prime} T}]}}_{T^\prime}]}}_R \Bigg)
		+ \dots 	\nonumber \\
	& = & H^{(0)}_{\lambda}+H^{(2)}_{\lambda}+H^{(3)}_{\lambda}+\dots ,
		\label{eh1}
\end{eqnarray}
where $H^B_{\lambda ij} = f_{\lambda ij} H^B_{ij}$ (we use the
notation $A_{ij} = \langle i | A | j \rangle$), $H^B_{I\lambda Tij}
= -{1\over E_j-E_i} \left({d\, \over d\lambda} f_{\lambda ij}\right)
H^B_{Iij} $, and
\begin{eqnarray}
	{\underbrace{X_{\lambda^\prime ij}}}_R &=& - f_{\lambda ij}
		\int_\lambda^\infty d\lambda^\prime X_{\lambda^\prime ij},
		\label{2} \\
	{\underbrace{X_{\lambda^\prime ij}}}_T &=& - {1 \over E_j - E_i}
		\Big({d\, \over d\lambda} f_{\lambda ij} \Big)
		\int_\lambda^\infty d\lambda^\prime X_{\lambda^\prime ij}
		+ {1 \over E_j -E_i} (1 - f_{\lambda ij}) X_{\lambda ij}.
		\label{3}
\end{eqnarray}
The function $f_{\lambda ij} = f(x_{\lambda ij})$ is a smearing
function in the similarity renormalization group, and
$x_{\lambda ij} = {E_j - E_i\over E_i + E_j + \lambda}$.  The smearing
function is introduced to force the  Hamiltonian $H_\lambda$
becoming a band diagonal form in energy space.  This requires
the following properties for $f_{\lambda ij}$: when $x < 1/3$, $f = 1$;
when $x > 2/3$, $f =0$; and $f$ may be a smooth function from
1 to 0 for $ 1/3 \leq x \leq 2/3$. Thus, through the similarity
renormalization group, we eliminate the interactions
between the states well-separated in energy and generate
the Hamiltonian of eq.(\ref{eh1}). The expansion of
eq.(\ref{eh1}) in terms of the interaction coupling
constant brings in order by order the full theory corrections
to this band diagonal low energy Hamiltonian.

Explicitly, the bare Hamiltonian $H^B$ input in the above formulation
can be obtained from the canonical Lagrangian with a high energy
cutoff that removes the usual UV divergences.  For light-front QCD
dynamics, the bare Hamiltonian in our consideration is the canonical
light-front QCD Hamiltonian that can be either obtained from the
canonical procedure in the light-front gauge \cite{Zhang93a,Zhang93b}
or generated from the light-front power counting rules \cite{Wilson94}.
Instead of the cutoff on the field operators which is introduced
in ref.\cite{Wilson94}, we shall use in this paper a vertex cutoff
to every vertex in the bare Hamiltonian:
\begin{equation}
	\theta(\Lambda^2/P^+ - |p_i^- - p_f^-|),   \label{ctf}
\end{equation}
where $p_i^-$ and $p_j^-$ are the initial and final state
light-front energies respectively between the vertex,
$\Lambda$ is the UV cutoff parameter, and $P^+$ the
total light-front longitudinal momentum of the system
we are interested in.  Eq.(\ref{ctf}) is also called the local
cutoff in light-front perturbative QCD\cite{Brodsky81}.
All the $\Lambda$-dependences in the final
bare Hamiltonian are removed by the counterterms
so that the bare Hamiltonian $H^B$ used in eq.(\ref{eh1}) has
already been renormalized as $\Lambda \rightarrow \infty$.
The use of eq.(\ref{ctf}) largely simplifies the analysis on
the cutoff scheme in ref.\cite{Wilson94}.

Meanwhile, in similarity renormalization group calculation, we should
also give an explicit form of the smearing function $f_{\lambda ij}$.
One of the simplest smearing functions that satisfies the requirements
of the similarity renormalization group scheme is a theta-function:
\begin{equation}
	f_{\lambda ij} = \theta ({1\over 2} - x_{\lambda ij}). \label{sm1}
\end{equation}
However, using the definition of $x_{\lambda ij}$, we can further replace
the above smearing function by the following form on the light-front:
\begin{equation}
	f_{\lambda ij} = \theta({\lambda^2 \over P^+} - |\Delta P_{ij}^-| ),
		\label{sm2}
\end{equation}
where  $\lambda$ is a low energy cutoff, and $\Delta P_{ij}^- =
P_i^- - P_j^-$
is the light-front free energy difference between the initial and final
states of the physical processes.  The light-front free energies of the
initial and final states are defined as sums over the light-front
free energies of the constituents in the states. The smearing function
eq.(\ref{sm2}) satisfies the requirements for the similarity
renormalization group approach although it is not a smooth function.

Throughout this paper, we shall always use the definition of
eq.(\ref{sm2}).  Thus, the  Hamiltonian (\ref{eh1})
can be reduced to
\begin{eqnarray}
	H_{\lambda ij} &=& \theta({\lambda^2 / P^+} - |\Delta
		P_{ij}^-| ) \left\{ H_{ij}^B + \sum_k H^B_{Iik}
		H^B_{Ikj} \Big[ \frac{g_{\lambda jik}}{\Delta P^-_{ik}}
		+ \frac{g_{\lambda ijk}}{\Delta P^-_{jk}} \Big]
		+ \cdots \right\} .  \label{eh2}
\end{eqnarray}
The front factor
(the theta-function) in the above equation indicates that the
 Hamiltonian $H_\lambda$ describes the low energy interactions
(with respect to the cutoff $\lambda$).  Therefore, $\lambda$ should
be a value of the hadronic mass scale.  Eq.(\ref{eh2}) shows that
the low energy Hamiltonian is apparently a modified
Hamiltonian perturbative expansion. The function $g_{\lambda ij}$
in eq.(\ref{eh2}) is
\begin{equation}
	g_{\lambda ijk} = \int_{\lambda^2/P^+}^\infty d({\lambda^2/ P^+})
		f_{\lambda' ik} {d \over d({\lambda'}^2/P^+)}
		f_{\lambda' jk} = \theta(|\Delta P_{jk}^-|-{\lambda^2/P^+})
		\theta( |\Delta P_{jk}^-| - |\Delta P_{ik}^-|) .
		\label{sd}
\end{equation}
The theta function in eq.(\ref{sd}) guarantees that the
singularity coming from the small energy denominators in the
usual Hamiltonian perturbation theory does not occur
in the above formulation.

Up to this point, we have introduced two cutoffs,
$\Lambda$ and $\lambda$.  The cutoff $\Lambda$ is used to remove
the UV divergences as in the usual perturbation theory.
While the low energy cutoff $\lambda$ is introduced to
separate the high and low energy state interactions, it is
indeed a scale parameter described by the
 light-front similarity renormalization group.
The similarity renormalization group transformations integrate
out all physical degrees of freedom above the cutoff $\lambda$
and generate the low energy Hamiltonian $H_\lambda$.  Formally,
when $\lambda \rightarrow \infty$, $f_{\lambda ij} = 1$,
and eqs.(\ref{2}) and (\ref{3}) vanish so that $H_\lambda
\rightarrow H^B$.  In practice, we shall take $\lambda
\sim 1$ GeV. Of course, the final physical observables
must be $\lambda$ independent.   As a
consequence of this condition, the coupling
constant and the constituent masses in the
Hamiltonian $H_\lambda$ become functions
of this low energy cutoff $\lambda$.  It is these
cutoff (or scale) dependences that link the effective
theory to the full theory in the limit $g \rightarrow g_s$.

Next, to calculate explicitly the low energy effective QCD
Hamiltonian, we begin with the canonical QCD theory.
It is convenient to use the light-front two-component
formulation of canonical QCD in the light-front gauge
$A_a^+=0$. The bare light-front QCD Hamiltonian
in such a two-component formalism is given by \cite{Zhang93b}
\begin{equation}
	H^B = \int_c dx^+ d^2 x_{\bot} \Big( {\cal H}_0 +
		{\cal H}_I \Big) . \label{qcdh}
\end{equation}
Here $\int_c$ means that the local cutoff, eq.(\ref{ctf}),
has been imposed, and
\begin{eqnarray}
	{\cal H}_0 &=& {1\over 2} (\partial^i A^j_a)(\partial^i A^j_a)
		+ \xi^{\dagger} {-\partial^2_{\bot} + m^2 \over
		i\partial^+} \xi ,  \\
	{\cal H}_I &=& {\cal H}_{qqg} + {\cal H}_{ggg} + {\cal H}_{qqqq}
		+{\cal H}_{qqgg} + {\cal H}_{gggg} + {\rm counterterms},
		\label{hdqcd}
\end{eqnarray}
with
\begin{eqnarray}
	& & {\cal H}_{qqg} = g \xi^{\dagger} \left\{ - 2 \left(
		\frac{1}{\partial^+} \right) ( \partial \cdot
		A_{\bot}) + \tilde{\sigma} \cdot A_{\bot} \left(
		\frac{1}{\partial^+} \right) (\tilde{\sigma}
		\cdot \partial_{\bot} + m) \right. \nonumber \\
	& & ~~~~~~~~~~~~~~~~~~~~~~ + \left. \left( \frac{1}{\partial^+}
		\right) (\tilde{\sigma} \cdot \partial_{\bot} - m)
		\tilde{\sigma} \cdot A_{\bot} \right\} \xi , \label{qqg} \\
	& & {\cal H}_{ggg} = g f^{abc} \left\{ \partial^i A_a^j A_b^i A_c^j
		+ (\partial^i A_a^i) \left( \frac{1}{\partial^+} \right)
		(A_b^j \partial^+ A_c^j) \right\} , \\
	& & {\cal H}_{qqgg} = g^2 \left\{ \xi^{\dagger} \tilde{\sigma}
		\cdot A_{\bot} \left( \frac{1}{i \partial^+} \right)
		\tilde{\sigma} \cdot A_{\bot} \xi \right. \nonumber \\
	& & ~~~~~~~~~~~~~~ \left. +~2  \left(\frac{1}{\partial^+}
		\right)(f^{abc} A_b^i \partial^+ A_c^i) \left( \frac{1}{
		\partial^+} \right) (\xi^{\dagger} T^a \xi) \right\} , \\
	& & {\cal H}_{qqqq} = 2g^2 \left\{ \left(\frac{1}{\partial^+}
		\right)(\xi^{\dagger} T^a \xi ) \left( \frac{1}
		{\partial^+} \right) (\xi^{\dagger} T^a \xi) \right\} , \\
	& & {\cal H}_{gggg} = \left. \frac{g^2}{4} f^{abc} f^{ade}
		\right\{ A_b^i A_c^j A_d^i A_e^j \nonumber \\
	& & ~~~~~~~~~~~~~~~~~ \left. + 2 \left(\frac{1}{\partial^+}
		\right)(A_b^i \partial^+ A_c^i) \left( \frac{1}{\partial^+}
		\right) (A_d^j \partial^+ A_e^j) \right\},  \label{gggg}
\end{eqnarray}
where $A_{\bot}=T^a A_{a\bot}$ is the transverse component of the
 gauge field, $T^a$ is the generator of SU(3) color group, and $\xi$ is
the two-component form of the light-front quark field:
\begin{equation}
	\psi_+ = \Lambda^+ \psi = \left[ \begin{array}{l} \xi \\
		0 \end{array} \right]~~,~~
	\psi_- =  \Lambda^- \psi =\left[ \begin{array}{c} 0 \\
		\left(\frac{1}{i\partial^+} \right)[\tilde{\sigma}^i
		(i \partial^i + gA^i) + im] \xi \end{array} \right] .
\end{equation}
The notation $\tilde{\sigma}$ is defined by: $\tilde{\sigma}^1 =
\sigma^2, \tilde{\sigma}^2 = - \sigma^1$ (the Pauli matrices).
This comes from the use of the light-front $\gamma$-representation:
\begin{eqnarray}
	&& \gamma^0 = \left[\begin{array}{cc} 0 & -i \\
		i & 0 \end{array} \right]~~, ~~~~
	   \gamma^3 = \left[\begin{array}{cc} 0 & i \\
		i & 0 \end{array} \right] , \nonumber \\
	&& \gamma^1 = \left[\begin{array}{cc} -i \sigma^2 & 0 \\
		0 & i \sigma^2 \end{array} \right]~~, ~~~~
	   \gamma^2 = \left[\begin{array}{cc} i \sigma^1 & 0 \\
		0 & -i\sigma^1 \end{array} \right], \label{lfgm}
\end{eqnarray}
which is different from the one we used in ref.\cite{Zhang93b}.
The reason of using this new $\gamma$-representation is that
it leads to not only a realization of the two-component form
for the light-front fermion field, but also a correct correspondence
of the fermion spin operator $S^i \sim \sigma^i$ on the light-front.
Counterterms are added to eq.(\ref{hdqcd}) in order to remove
all the $\Lambda$-dependence. Thus the coupling constant $g$
in eqs.(\ref{qqg}--\ref{gggg}) is a perturbatively
renormalized coupling constant.

Upon the calculation to the second order in the coupling
constant, we obtain an effective Hamiltonian (\ref{eh2}) in
$q\overline{q}$ sector (with the initial and final states,
$| i \rangle = b^{\dagger}(p_1, \lambda_1)
d^{\dagger}(p_2,\lambda_2) | 0 \rangle$ and $| j \rangle =
b^{\dagger}(p_3, \lambda_3) d^{\dagger}(p_4,\lambda_4) | 0
\rangle$, respectively, where $p_i$ and $\lambda_i$ denote
the respective momentum and  helicity of a quark on the light-front):
\begin{eqnarray}
	H_{\lambda{ij}} &=& \theta({\lambda^2 / P^+} - |\Delta
		P_{ij}^-| ) \Bigg\{ \langle j |:H^B:| i \rangle
		- {g^2 \over 2\pi^2}\lambda^2 C_f {1 \over P^+}
		\ln \epsilon + {\rm mass ~ counterterms}
		\nonumber \\
	& & ~~~~~~~~~~~~~ -g^2 (T^a)(T^a) {\theta(q^+) \over q^+}
		\chi^{\dagger}_{\lambda_1} \left[2{ q^{i'}_{\bot}
		\over q^+} - {\tilde{\sigma} \cdot p_{3\bot}-im
		\over [p_3^+]} \tilde{\sigma}^{i'} - \tilde{\sigma}^{i'}
		{\tilde{\sigma} \cdot p_{1\bot} + im
		\over [p_1^+]} \right] \chi_{\lambda_3}\nonumber \\
	& & ~~~~~~~~~~~~~~~~~~~~~~\times \chi^{\dagger}_{-\lambda_2}
		\Bigg[2{ q^{i'}_{\bot} \over q^+} - {\tilde{\sigma}
		\cdot p_{2\bot} + im \over [p_2^+]} \tilde{\sigma}^{i'}
		- \tilde{\sigma}^{i'} {\tilde{\sigma} \cdot
		p_{4\bot} - im \over [p_4^+]} \Bigg] \chi_{-\lambda_4}
		F_{rij} \Bigg\} . \label{eh}
\end{eqnarray}
Here $P^+= p_1^+ + p_2^+ = p_3^+ + p_4^+$, $\Delta P_{ij}^- = p_1^-
+ p_2^- - p_3^- - p_4^-$, $:H^B:$ represents a normal ordering,
where the instantaneous interaction contribution to the quark
self-energy has been included in the self-energy calculation which is
given by the mass counterterms and the logarithmic divergence in
(\ref{eh}), the color factor $C_f=(T^a T^a) = (N^2-1)/2N$, $N=3$
the total numbers of colors, and $\epsilon$ is an infrared longitudinal
momentum cutoff.  Since $\ln \epsilon$ is an infrared
divergence, it cannot be removed by mass counterterms.
In  gauge symmetry, this divergence must be canceled in the
physical sector (and this is true as we will see later).
The last term in (\ref{eh}) is the one-gluon exchange contribution
to the low energy Hamiltonian.  The momentum $q$ is carried by
the exchange gluon: $q^+ = p_1^+ - p_3^+ = p_4^+ - p_2^+$,
$q_{\bot} = p_{1\bot} - p_{3\bot} = p_{4\bot} - p_{2\bot}$,
$\chi_{\lambda_i}$ denotes a helicity eigenstate.
The factor $F_{rij}$ arises from the similarity renormalization
group scheme:
\begin{eqnarray}
	F_{rij} &=& \Bigg\{\theta( {\Lambda^2/P^+} - |p_1^- - p_3^-
		-q^-|) \theta( {\Lambda^2/P^+} - |p_4^- - p_2^- - q^-|)
		\nonumber \\
	& & ~~ \times \Bigg[{\theta(|p_1^- - p_3^- -q^-|
		- {\lambda^2 \over P^+}) \theta( |p_1^- - p_3^- -q^-| - |p_4^-
		- p_2^- - q^-|) \over p_1^- - p_3^- -q^- } \nonumber \\
	& &  ~~~~~~~ + {\theta(|p_4^- - p_2^- - q^-| - {\lambda^2 \over
		P^+}) \theta(|p_4^- - p_2^- - q^-| - |p_1^- - p_3^- -q^-|)
		\over p_4^- - p_2^- - q^-} \Bigg] \nonumber \\
	& & ~~~~~~~ + (p_1 \longleftrightarrow p_3~, ~~ p_2
		\longleftrightarrow p_4~, ~~ q \longrightarrow -q)
		\Bigg\}.  \label{fcsr}
\end{eqnarray}
All light-front energies in eq.(\ref{fcsr}) are on mass-shell:
$p_i^- = {p_{i\bot}^2 + m_i^2 \over p_i^+}$ and $q^- =
{q_{\bot}^2 \over q^+}$.

If we continue to evaluate all the terms in the expansion of
eq.(\ref{eh1}), the resulting Hamiltonian is the exact
QCD Hamiltonian. In practice, we should only consider the
leading and the next-to-leading terms, i.e., eq.(\ref{eh}), as a
starting effective Hamiltonian. The basic idea to realize
a weak-coupling treatment of QCD for hadrons is whether we can
solve hadron states from this effective Hamiltonian (\ref{eh})
with an arbitrary small coupling constant $g$ such
that the higher order corrections in (\ref{eh1}) can be handled
perturbatively. From the success of the constituent quark model,
we understand that a necessity for such a realization is the
existence of a confining interaction in (\ref{eh}).

Naively, we know that any weak-coupling Hamiltonian
derived from QCD will have only Coulomb-like interactions, and
confinement can only be exhibited in a strong-coupling theory.
However, Perry has recently found that upon to the
second order calculation of the low energy Hamiltonian,
a logarithmic confining potential has already occurred
\cite{Perry94}.  Explicitly, the two-body
quark-antiquark interaction from the first term of
eq.(\ref{eh}) is the instantaneous gluon exchange
interaction ${\cal H}_{qqqq}$ which has the form in
the momentum space:
\begin{equation}
	- {1 \over (q^+)^2} .
\end{equation}
The confinement must be associated with the interaction where
$q^+ \rightarrow 0$.  This is because only these particles with
zero longitudinal momentum can occupy the
light-front vacuum state\cite{Wilson94}.  Thus, we need
only to analyze the feature of the effective Hamiltonian
when $q^+ \rightarrow 0$.  For $q^+ \rightarrow 0$, the dominant
second order contribution from one transverse gluon
exchange interaction in eq.(\ref{eh}) is given by
\begin{equation}	\label{ogx}
    \Bigg[{1 \over (q^+)^2} {q^2_{\bot} \over q^2_{\bot} } +
	O({1 \over q^+})\Bigg] \theta (q^- - \lambda^2/P^+) .
\end{equation}
In the usual perturbative calculation, no such a $\theta$-function
is attached in the above equation. Thus these two dominant
contributions from the instantaneous and one-gluon exchange
interactions are exactly cancelled when $q^+ \rightarrow 0$.  Only
Coulomb-type interaction (the terms $\sim O({1\over q^+})$) remains.
However, in the light-front similarity renormalization group scheme,
the one-gluon exchange contribution to eq.(\ref{ogx})
only contains these gluons
with energy being greater than the energy cutoff $\lambda^2 /P^+$.
As a result, the instantaneous gluon exchange term $1/(q^+)^2$
remains uncancelled if the gluon energy, $q^2_{\bot} /q^+$, is
less than $\lambda^2 /P^+$.  The remaining uncancelled instantaneous
interaction contains an infrared divergence and a finite part
contribution (for a detailed derivation, see section V).
The divergence part is cancelled precisely for physical states
by the same divergence in the quark self-energy correction
[see (\ref{eh})]. The remaining
finite part corresponds to a logarithmic confining potential:
\[  b_{\lambda} \ln |x^-| + c_{\lambda} \ln\Big( {\lambda^2
	|x_{\bot}|^2 \over P^+}\Big) . \]
The above result is first obtained by Perry with the use of the concept
of coupling coherence and a slightly different renormalization
scheme \cite{Perry94} (also see a oversimple derivation given
by Wilson \cite{Wilson94a}).  Here the derivation is purely
based on the light-front similarity renormalization scheme
\cite{Wilson94}.

One may argue that the existence of such a confining potential
in $H_{\lambda}$ may only be an artificial effect
designed in the renormalization scheme we used.  If we included
the interaction with the exchange gluon energy below the cutoff,
then the instantaneous interaction would be completely cancelled,
and no such confining potential should exist, as expected in
the usual perturbation computation. Wilson has pointed out that the
set up of the new renormalization scheme is motivated by the idea
that the gluon mass must be nonzero in the low energy domain
(a constituent picture), which could be regarded
as an effect of the nontrivial low energy gluon
interactions. The cutoff $\lambda$ is of the same order
as the constituent gluon mass. Thus the gluon energy cannot
run down below the cutoff $\lambda$.
The existence of the confining potential is a result
of the low-energy gluon interactions. In contrast, the
photon mass in QED is zero at any scale. The
instantaneous photon exchange interaction is always cancelled
by the corresponding one transverse photon exchange
interaction, and only the Coulomb interaction is left. Thus,
the above confining potential can only exist in QCD
\cite{Wilson94a}. More detailed discussions will be seen
in the subsequent sections.

Yet,  even up to the second order, the effective QCD
Hamiltonian $H_{\lambda}$ is already very complicated.
In order to to examine the above ideas of confining mechanism
and to develop explicitly a weak-coupling treatment approach of
nonperturbative QCD to hadronic bound states, in the next
section we shall utilize the above formulation to heavy
quark systems. We find that the low-energy QCD Hamiltonian
$H_{\lambda}$ for heavy quarks can be largely simplified and
an analytic form consisting of the confining potential
and Coulomb potential emerges.

\section{Low-energy heavy quark confining Hamiltonian}

In the past few years, QCD has been made a numerous progresses in
understanding the heavy hadron structure, due mainly to the discovery
of heavy quark symmetry by Isgur and Wise \cite{Isgur90}, and the
development of the heavy quark effective theory by Georgi et al.
\cite{Georgi}.
Very recently, we have reformulated the heavy quark effective
theory from QCD on the light-front \cite{Zhang95,Cheung95}, which
may provide a convenient basis for the further study of the
nonperturbative structures of heavy hadron. Now we use the
light-front HQET to derive
the low energy heavy quark effective Hamiltonian in the similarity
renormalization group scheme.

\subsection{Light-front heavy quark effective theory}

The light-front heavy quark effective Lagrangian derived from
QCD lagrangian ${\cal L} = \overline{Q}(i {\not \! \! D}
-m_Q) Q$ as a $1/m_Q$ expansion is given in
refs.\cite{Zhang95,Cheung95}:
\begin{eqnarray}
       && {\cal L} = {2 \over v^+} {\cal Q}_{v+}^{\dagger} (iv \cdot D)
                {\cal Q}_{v+} - \sum_{n=1}^{\infty} \Big({ 1 \over m_Q v^+}
                \Big)^n {\cal Q}_{v+}^{\dagger} \Big\{(i\vec{\alpha}
                \cdot \vec{D})
	(-i D^+)^{n-1}(i\vec{\alpha} \cdot \vec{D}) \Big\} {\cal Q}_{v+},
\end{eqnarray}
where ${\cal Q}_{v+}$ is the light-front dynamical component
of the heavy quark field after the phase redefinition:
\begin{equation}	\label{phase}
 	Q(x)= e^{-im_Q v \cdot x} ({\cal Q}_{v+}(x) + {\cal Q}_{v-}(x)),
\end{equation}
$v^{\mu}$ the four velocity of the heavy hadrons, $P^\mu
=M_H v^\mu$ with $v^2=1$ and $M_H$ being the heavy hadron
mass, $ \vec{\alpha} \cdot \vec{D} \equiv \alpha_{\bot}
\cdot D_{\bot} - {1\over v^+} (\alpha_{\bot} \cdot v_{\bot} +
\beta) D^+$, and $D^{\mu}$ is the usual covariant derivative.
The corresponding light-front heavy quark bare Hamiltonian
density is given by
\begin{eqnarray}
    && {\cal H} = { 1\over iv^+} {\cal Q}^{\dagger}_{v+} (v^-\partial^+
                -2v_{\bot} \cdot \partial_{\bot} ) {\cal Q}_{v+}
                - {g \over v^+} {\cal Q}^{\dagger}_{v+} (v \cdot A)
                {\cal Q}_{v+} \nonumber \\
	 & & ~~~~~~~~~~~ + \sum_{n=1}^{\infty} \Big({ 1 \over m_Q v^+}
                \Big)^n {\cal Q}_{v+}^{\dagger} \Big\{(i\vec{\alpha}
                \cdot \vec{D}) (-i D^+)^{n-1} (i \vec{\alpha} \cdot
		\vec{D}) \Big\} {\cal Q}_{v+}.  \label{hqbeh}
\end{eqnarray}

In the large $m_Q$ limit, only the leading (spin and mass
independent) Hamiltonian is remained. In other words, the
phase redefinition (\ref{phase}) removes the dominant piece
of the space-time dependence of the heavy quark. The remaining
dependence is only due to the residual momentum of the heavy
quark in the heavy hadrons. The $1/m_Q^n$ terms ($n \geq 1$)
in (\ref{hqbeh}) can
be regarded as perturbative corrections to the leading order
operators and states. Therefore, the heavy quark mass
$m_Q$ is indeed a factorization scale for separating
heavy quark short and long distance dynamics. To determine
confining interactions between two heavy quarks or a heavy quark
with a light quark, only the leading heavy quark Hamiltonian
plays an essential role.
We choose the light-front gauge $A^+=0$, the leading-order
bare QCD Hamiltonian density (corresponding to the limit of $m_Q
\rightarrow \infty$) is given from (\ref{hqbeh}):
\begin{eqnarray}
    {\cal H}_{ld} &=& { 1\over iv^+} {\cal Q}^{\dagger}_{v+} (v^-\partial^+
                -2v_{\bot} \cdot \partial_{\bot} ) {\cal Q}_{v+} \nonumber \\
     & & ~~~~~~ - {2g \over v^+} {\cal Q}^{\dagger}_{v+} \left\{v^+ \Big[
		\Big({1\over \partial^+}\Big) \partial_{\bot} \cdot A_{\bot}
		\Big] - v_{\bot} \cdot A_{\bot} \right\} {\cal Q}_{v^+}
		\nonumber \\
     & & ~~~~~~ + 2g^2 \Big({1\over \partial^+}\Big) \Big({\cal
		Q}^{\dagger}_{v+} T^a {\cal Q}_{v^+} \Big) \Big({1 \over
		\partial^+} \Big) \Big( {\cal Q}^{\dagger}_{v+} T^a
		{\cal Q}_{v^+} \Big) \nonumber \\
     &= & {\cal H}_0 + {\cal H}_{qqg} + {\cal H}_{qqqq} . \label{ldhh}
\end{eqnarray}
Note that besides the leading term in eq.(\ref{hqbeh}),
the above bare Hamiltonian has  already also included
the relevant terms from the gauge field part, $-{1\over 2}
{\rm Tr} (F_{\mu\nu} F^{\mu\nu})$, of the QCD Lagrangian.
These terms come from the elimination of the
unphysical gauge degrees of freedom, the longitudinal
component $A^-_a$, in the light-front gauge
(see a detailed derivation in refs.\cite{Zhang93a,Zhang93b}).
Eq(\ref{ldhh}) has obviously the spin and flavour heavy
quark symmetry, or simply the heavy quark symmetry.

In momentum space, the free part of the bare light-front
effective heavy quark Hamiltonian can be simply expressed by
\begin{eqnarray}
	H_{0} &=& \sum_{\lambda} \int [d^3\bar{k}]
		{1 \over v^+} ( 2 v_{\bot} \cdot k_{\bot} - v^- k^+)
		\Big\{b^{\dagger}_v (k,\lambda) b_v(k,\lambda) +
		d^{\dagger}_v (k,\lambda) d_v(k,\lambda)\Big\} \nonumber \\
	&=& \sum_{\lambda} \int [d^3\bar{k}]
		k^- \Big\{b^{\dagger}_v (k,\lambda) b_v(k,\lambda) +
		d^{\dagger}_v (k,\lambda) d_v(k,\lambda)\Big\}	,
		\label{lffh}
\end{eqnarray}
where $k$ is the residual momentum of heavy quarks, $p^{\lambda}
= m_Q v^{\lambda} + k^{\lambda}$, and $\lambda$ its helicity.  We
have introduced the notation for the space components of light-front
momentum $ (p^+, p_\bot) \equiv \bar{p} $ so that $[d^3\bar{p}]
\equiv {dp^+ d^2 p_{\bot} \over 2(2\pi)^3}$.
The operator $b^{\dagger}_v(k,\lambda) ~[d^{\dagger}_v(k,
\lambda)]$ creates a heavy quark [antiquark]
with velocity $v$, residual momentum $k$ and helicity
$\lambda$,
\begin{equation}
	\{b_v(k,\lambda), ~ b^{\dagger}_{v'}(k',\lambda')\}
		=\{d_v(k,\lambda), ~ d^{\dagger}_{v'}(k',\lambda')\}
		= 2 (2\pi)^3 \delta_{vv'} \delta_{\lambda \lambda'}
		\delta^3 (\bar{k}-\bar{k}'),
\end{equation}
where $\delta^3(\bar{k}-\bar{k}') \equiv \delta(k^+-{k'}^+)
\delta^2(k_\bot-k'_\bot)$.

Eq.(\ref{lffh}) means that after the redefinition of the heavy
quark field, the heavy quark in the light-front HQET carries
the effective free light-front energy
\begin{equation}	\label{k-}
	k^- = {1\over v^+} ( 2 v_{\bot} \cdot k_{\bot}
		- v^- k^+) .
\end{equation}
The meaning of this result becomes more transparent
if we expand the light-front energy dispersion
relation of the heavy quarks as an inverse
power of $m_Q$:
\begin{eqnarray}
	p^- &&= {p_{\bot}^2 + m_Q^2 \over p^+} = m_Qv^- +
		{1\over v^+} ( 2 v_{\bot} \cdot k_{\bot}
		- v^- k^+) + O(1/m_Q) \nonumber \\
	&& \stackrel{m_Q \rightarrow \infty}{=} m_Qv^- + k^- .
		\label{pmk}
\end{eqnarray}
We see that the mass part $(m_Qv^-)$ has been removed by the phase
redefinition of (\ref{phase}). Thus, eq.(\ref{ldhh}) describes
effectively the ``lighten'' heavy quark dynamics
with respect to its residual momentum. In other words, the
heavy quarks in the effective theory have the same energy scaling
behavior as the light ones.

The above leading Hamiltonian (or Lagrangian) is the basis of the
QCD-based description for heavy hadrons containing a single heavy
quark, such as $B$ and $D$ mesons.  However, as recently pointed
out by Mannel et al.\cite{Mannel93,Mannel95} the purely heavy
quark leading Lagrangian may be not appropriate for the
description of heavy quarkonia states.  This is because the
anomalous dimension of the QCD radiative correction to
the $Q\overline{Q}$ currents contains an infrared singularity
in the limit of two heavy constituents having equal
velocity. Such an infrared singularity is a long distance
effect and should be absorbed into quarkonium states.
To avoid this problem, they argued that one may incorporate
the effective Hamiltonian with at least the first order kinetic
energy term into the leading Hamiltonian \cite{Mannel95}. The
kinetic energy in light-front HQET is given by \cite{Zhang95}
\begin{equation}
	{\cal H}_{kin} = - {1\over m_Q v^+} {\cal Q}_{v+}^\dagger
		\Bigg\{ \partial_\bot^2 - {2 v_\bot \cdot
		\partial_\bot \over v^+}
		\partial^+ + {v^- \over v^+} \partial^{+2} \Bigg\}
		{\cal Q}_{v+}.	\label{nloh}
\end{equation}
As a consequence, in the heavy mass limit, quarkonia have
spin symmetry but no flavour symmetry. In momentum space,
\begin{eqnarray}
	H_{kin} = {1 \over m_Qv^+} \sum_{\lambda} \int [d^3\bar{k}]
		\Big( k_\bot^2 && - 2 v_{\bot} \cdot k_{\bot}{k^+
		\over v^+} + {v^- \over v^+} k^{+2} \Big) \nonumber \\
	& & \times \Big\{b^{\dagger}_v (k,\lambda) b_v(k,\lambda) +
	  d^{\dagger}_v (k,\lambda) d_v(k,\lambda)\Big\}. \label{hke}
\end{eqnarray}
The kinetic energy of (\ref{hke}) can be simply obtained
by expanding (\ref{pmk}) up to order $1/m_Q$.
We will discuss later the effect of this kinetic energy in
the determination of heavy quarkonium bound states.

\subsection{Low-energy effective Hamiltonian for heavy quarkonia}

Within light-front HQET, we can follow the procedure described in
the previous section to find the effective low energy  QCD
Hamiltonian for $Q\overline{Q}$ systems. The bare Hamiltonian
for $Q\overline{Q}$ systems is given by (\ref{ldhh}) plus
the leading kinetic Hamiltonian (\ref{nloh}) for both
heavy quark and antiquark, where two heavy constituents have
the same velocity carried by the heavy quarkonia. The kinetic
energy which is of order $\Lambda_{QCD}/m_Q$, is at most
the same order as the Coulomb interaction. It may affect on
the quarkonium bound states but not on the
derivation of the long distance quark interactions. In fact,
the success of the potential-model description
indicates that the scalar interactions between the two heavy
constituents are flavour-independent. In other words, the
confining interaction in $Q\overline{Q}$ states should
be independent of the kinetic energy (\ref{nloh}).
Therefore, we may treat the kinetic energy as the same
as the instantaneous $Q\overline{Q}$ interaction
[the last term in eq.(\ref{ldhh})].  In the derivation
of the low energy heavy quark Hamiltonian, the free Hamiltonian
used in similarity renormalization group scheme is then
simply given by eq.(\ref{lffh}).

With the above consideration,
it is easy to find that the leading order contribution
to the low energy effective Hamiltonian is the low energy
part of the  heavy quark effective bare Hamiltonian,
\begin{eqnarray}
    	H_{\lambda ij}^{(0)}= \theta({\lambda^2\over P^+} -|\Delta
		P_{ij}^-|) \langle j |
		\int dx^- && d^2x_{\bot} \Bigg\{  { 1\over iv^+}
		{\cal Q}^{\dagger}_{v+} (v^-\partial^+ -2v_{\bot}
		\cdot \partial_{\bot} ) {\cal Q}_{v+} \nonumber \\
     & & - {2g \over v^+} {\cal Q}^{\dagger}_{v+}
		\left\{v^+ \Big[\Big({1\over \partial^+}\Big)
		\partial_{\bot} \cdot A_{\bot}\Big] - v_{\bot}
		\cdot A_{\bot} \right\} {\cal Q}_{v^+} \nonumber \\
     & & + 2g^2 \Big({1\over \partial^+}\Big) \Big({\cal
		Q}^{\dagger}_{v+} T^a {\cal Q}_{v^+} \Big) \Big({1 \over
		\partial^+} \Big) \Big( {\cal Q}^{\dagger}_{v+} T^a
		{\cal Q}_{v^+} \Big)  \nonumber \\
     & &  - {1\over m_Q v^+} {\cal Q}_{v+}^\dagger \Big[\partial_\bot^2
		- {2 v_\bot \cdot \partial_\bot \over v^+}
		\partial^+ + {v^- \over v^+} \partial^{+2} \Big]
		{\cal Q}_{v+} \Bigg\} | i \rangle, \label{lll}
\end{eqnarray}
plus all other $1/m_Q$ terms in (\ref{hqbeh}) as well as the
light quark and gluon full QCD Hamiltonian that has not been
included in the above equation [see (\ref{qcdh})].
Here the initial and final states are defined by $| i \rangle =
b^{\dagger}_v(k_1, \lambda_1) d^{\dagger}_v(k_2,\lambda_2)
| 0 \rangle$ and $| j \rangle = b^{\dagger}_v(k_3, \lambda_3)
d^{\dagger}_v(k_4,\lambda_4) | 0 \rangle$, respectively,
$P^+=p_1^++p_2^+ = (m_Q + m_{\overline{Q}})v^+ +k_1^+ + k_2^+
= (m_Q + m_{\overline{Q}})v^+ +k_1^+ + k_2^+$.

The next-to-leading order contribution contains two different
parts, $H_\lambda^{(2)} = H_{\lambda 1}^{(2)} + H_{\lambda 2}^{(2)}$,
where $ H_{\lambda 1}^{(2)}$ is the self-energy correction,
\begin{eqnarray}
	 H_{\lambda 1}^{(2)}&&=\theta({\lambda^2\over P^+}
		-  |\Delta P_{ij}^-|) [2(2\pi)^3]^2 \delta^3
		(\bar{k}_1-\bar{k}_3) \delta^3(\bar{k}_2-\bar{k}_4)
		\delta_{\lambda_1 \lambda_3} \delta_{\lambda_2 \lambda_4}
		(4g^2) (T^aT^a) \nonumber \\
	&& ~~~~\times \int [d^3\bar{k}] \Bigg\{ {\theta(k_1^+-k^+)
		\over k_1^+-k^+} \Bigg({ (k_1-k)^{i'}_{\bot}
		\over k_1^+-k^+} - {v^{i'}\over v^+} \Bigg)^2
		\theta({\Lambda^2 \over P^+} - |k_1^-
		- k^- -(k_1-k)^-|) \nonumber \\
	&&~~~~~~~~~~~~~~~~~~~~~~~~~~~~~\times {\theta(|k_1^- - k^-
		- (k_1-k)^-| - {\lambda^2 \over P^+}) \over k_1^- - k^-
		-(k_1-k)^- } \nonumber \\
	&& ~~~~~~~~~~~~~~ + {\theta(k_2^+-k^+) \over k_2^+-k^+}
		\Bigg({ (k_2-k)^{i'}_{\bot} \over k_2^+-k^+}
		- {v^{i'}\over v^+} \Bigg)^2 \theta( {\Lambda^2
		\over P^+} - |k_2^- - k^- -(k_2-k)^-|) \nonumber \\
	&& ~~~~~~~~~~~~~~~~~~~~~~~~~~~ \times {\theta(|k_2^- - k^-
		- (k_2-k)^-| - {\lambda^2 \over P^+}) \over k_2^- - k^-
		-(k_2-k)^- } \Bigg\} \nonumber \\
	&& = \theta({\lambda^2\over P^+} - |\Delta P_{ij}^-|)
		[2(2\pi)^3]^2 \delta^3 (\bar{k}_1-\bar{k}_3)
		\delta^3(\bar{k}_2-\bar{k}_4) \delta_{\lambda_1
		\lambda_3} \delta_{\lambda_2 \lambda_4} \nonumber \\
	&& ~~~~~~~~~~~~~~~ \times {-8g^2 C_f \over P^+}
		\int {dx_1 d^2\kappa_{1\bot} \over
		2(2\pi)^3} {\theta(x-x_1)\over (x-x_1)^2}
		F(x-x_1, \kappa_\bot - \kappa_{1\bot},M_H) \nonumber \\
	&& ~~~~~~~~~~~~~~~~~~~~~~~~~~~~~~~~~~~~~~~ \times
		{(\kappa_{\bot}-\kappa_{1\bot})^2 \over (\kappa_{\bot}
		-\kappa_{1\bot})^2 + (x-x_1)^2 M_H^2} ,  \label{sf}
\end{eqnarray}
and $H_{\lambda 2}^{(2)}$ is the $Q\overline{Q}$ interaction,
\begin{eqnarray}
	 H_{\lambda 2}^{(2)} && = \theta({\lambda^2\over P^+}
		-|\Delta P_{ij}^-|) \delta_{\lambda_1 \lambda_3}
		\delta_{\lambda_2 \lambda_4} 2(2\pi)^3 \delta^3
		(\bar{k}_1 + \bar{k}_2 - \bar{k}_3 - \bar{k}_4)
		 (-4g^2)(T^a)(T^a) \nonumber \\
	& & ~~~~~~~~\times  { 1 \over q^+} \Bigg({ q^{i'}_{\bot}
		\over q^+} - {v^{i'}\over v^+} \Bigg)^2 \theta({\Lambda^2
		\over P^+} - |k_1^- - k_3^- - q^-|) \theta({\Lambda^2
		\over P^+} - |k_4^- - k_2^- - q^-|) \nonumber \\
	& & ~~~~~~~~~~~~~\times \Bigg\{{\theta(|k_1^- - k_3^- - q^-| -
		{\lambda^2 \over P^+})\theta(|k_1^- - k_3^- - q^-| -
		|k_4^- - k_2^- - q^-|) \over k_1^- - k_3^- - q^-}
			 \nonumber \\
	& & ~~~~~~~~~~~~~~~~+ {\theta(|k_4^- - k_2^- - q^-| - {\lambda^2
		\over P^+}) \theta(|k_4^- - k_2^- - q^-| - |k_1^- -
		k_3^- - q^-|) \over k_4^- - k_2^- - q^-} \Bigg\}
		\nonumber \\
	& & = \theta({\lambda^2\over P^+} -|\Delta P_{ij}^-|) 2(2\pi)^3
		\delta^3(\bar{k}_1 + \bar{k}_2 - \bar{k}_3 - \bar{k}_4)
		\delta_{\lambda_1 \lambda_3} \delta_{\lambda_2 \lambda_4}
		F(x-x', \kappa_\bot - \kappa'_\bot,M_H)  \nonumber \\
	& & ~~~~~~~~~~~~~~~~~~~~ \times { 4g^2
		(T^a)(T^a) \over (P^+)^2} { 1 \over (x-x')^2}
		{(\kappa_{\bot}-\kappa_{\bot}')^2 \over
		(\kappa_{\bot}-\kappa_{\bot}')^2 + (x-x')^2
		M_H^2} ,  \label{vv}
\end{eqnarray}
where $k^-$ is given by eq.(\ref{k-}), $q^+ = k_1^+ - k_3^+ =
k_4^+ - k_2^+$, $q_\bot = k_{1\bot} -
k_{3\bot} = k_{4\bot} - k_{2\bot}$, and $q^- = q_\bot^2 /q^+$,
we have also introduced the longitudinal residual momentum
fractions and the relative transverse residual momenta,
\begin{eqnarray}
	&& x = k^+_1 / P^+ ~,~~~~~ \kappa_{\bot} =
		k_{1\bot} - x P_{\bot} , \nonumber \\
	&& x' = k^+_3 / P^+ ~,~~~~~ \kappa'_{\bot} =
		k_{3\bot} - x' P_{\bot},
\end{eqnarray}
and defined the function $F$,
\begin{equation}
	F(x, k, M) \equiv \theta (A(x, k, M) - \lambda^2) \theta(
		\Lambda^2 - A(x,k,M)),
\end{equation}
with
\begin{equation}
	A(x,k,M) \equiv {k^2 \over |x|} + |x| M^2 .
\end{equation}
Since  $ 0 \leq p_1^+ = m_Q v^+ + k_1^+ \leq P^+=M_Hv^+$,
in the heavy quark mass limit, we have $M_H \rightarrow 2m_Q$
so that $-m_Qv^+ \leq k^+_1, ~ k^+_3 \leq m_Q v^+$.  Hence,
the range of $x$ and $x'$ is given by
\begin{equation}
	 -{1 \over 2} \leq x, ~ x' \leq {1 \over 2} .
\end{equation}
Eqs.(\ref{lll}), (\ref{sf}) and (\ref{vv}) consist of the effective
Hamiltonian for quarkonia up to the second order in
the similarity renormalization group scheme.

Apparently, the above effective Hamiltonian
is not a low energy Hamiltonian because $P^+=M_H v^+$
which is of order a few GeV.  However, from eq.(\ref{lffh})
and eq.(\ref{pmk}) we see that in the bare heavy quark
Hamiltonian the mass term $m_Q v^-$ has been integrated
out.  To address the correct energy scale of the low energy
heavy hadron dynamics, we should introduce the residual center
mass momentum of the heavy quarkonia $K^\mu$
and the residual heavy hadron mass $\overline{\Lambda}$,
\begin{equation}
	K^\mu = \overline{\Lambda} v^{\mu}$~~, ~~~ $\overline{\Lambda}
		= M_H - m_{Q} - m_{\overline{Q}},
\end{equation}
where $v^\mu$ is still the four-velocity of hadrons. It follows
that
\begin{equation}
	K^+ = k_1^+ + k_2^+ = k_3^+ + k_4^+ ~~,
		~~~ K_{\bot}= k_{1\bot} + k_{2\bot} = k_{3\bot}
		+ k_{4\bot}.
\end{equation}
With the residual heavy hadron momentum $K^\mu$ considered,
eqs.(\ref{sf}) and (\ref{vv}) become
\begin{eqnarray}
	 H_{\lambda 1}^{(2)} = \theta({\lambda^2\over K^+}
		- && |\Delta K_{ij}^-|) [2(2\pi)^3]^2 \delta^3
		(\bar{k}_1-\bar{k}_3) \delta^3(\bar{k}_2-\bar{k}_4)
	 	\delta_{\lambda_1 \lambda_3} \delta_{\lambda_2 \lambda_4}
			\nonumber \\
	&&  \times {-8g^2 C_f \over K^+ }\int {dy_1 d^2\kappa_{1\bot} \over
		2(2\pi)^3} {\theta(y-y_1)\over (y-y_1)^2} F(y-y_1,
		\kappa_\bot - \kappa_{1\bot},\overline{\Lambda})
		\nonumber \\
	&& ~~~~~~~~~~~~~~~~~~~~~~~~~~~~~~~ \times  {(\kappa_{\bot}
		-\kappa_{1\bot})^2 \over (\kappa_{\bot}
		- \kappa_{1\bot})^2 + (y-y_1)^2 \overline{\Lambda}^2} ,
		\label{sfl}
\end{eqnarray}
\begin{eqnarray}
	 H_{\lambda 2}^{(2)} = \theta({\lambda^2\over K^+}
		- && |\Delta K_{ij}^-|)  2(2\pi)^3 \delta^3(\bar{k}_1
		+ \bar{k}_2 - \bar{k}_3 - \bar{k}_4) \delta_{\lambda_1
		\lambda_3} \delta_{\lambda_2 \lambda_4} F(y-y',
		\kappa_\bot - \kappa'_\bot,\overline{\Lambda}^2)
		\nonumber \\
	& & ~~~ \times {4g^2 (T^a)(T^a)
		\over (K^+)^2} { 1 \over (y-y')^2}
		{(\kappa_{\bot}-\kappa_{\bot}')^2 \over
		(\kappa_{\bot}-\kappa_{\bot}')^2 + (y-y')^2
		\overline{\Lambda}^2}, \label{vvl}
\end{eqnarray}
where we have also introduced the corresponding residual
relative momenta:
\begin{eqnarray}
	&& y = k^+_1 / K^+ ~,~~~~~ \kappa_{\bot} =
		k_{1\bot} - y K_{\bot} , \nonumber \\
	&& y' = k^+_3 / K^+ ~,~~~~~ \kappa'_{\bot} =
		k_{3\bot} - y' K_{\bot}.	\label{QQlmf}
\end{eqnarray}
The range of the residual longitudinal momentum
fractions $y$ and $y'$ are given by
\begin{equation}
	- \infty < y = {M_H \over \overline{\Lambda}} x < \infty~~,
	~~ - \infty < y' = {M_H \over \overline{\Lambda}} x' < \infty .
\end{equation}
Now all the quantities appearing in the effective
Hamiltonian  have the low energy scale of
a few MeV.

\subsection{Reexpression of the low energy Hamiltonian in the
weak-coupling treatment scheme}

As we have mentioned in the Introduction, the first step to
follow the idea of the weak-coupling approach is to construct
the low energy Hamiltonian $H_\lambda$ up to the second order
and then separate it into $H_{\lambda 0}$ and $H_{\lambda I}$,
\begin{equation}
	H_{\lambda} = H_{\lambda 0} + H_{\lambda I}.  \label{seh}
\end{equation}
In eq.(\ref{seh}), $H_{\lambda 0}$ contains the free
Hamiltonian plus the dominant
two-body interactions which conserve the particle number, and
$H_{\lambda I}$ is the remaining interaction Hamiltonian
which describes the emission and reabsorption processes plus
all the higher order terms in the expansion of eq.(\ref{eh1}).
{\it Once $H_\lambda$ is derived, we can reexpress it as
eq.(\ref{seh}) such that $H_{\lambda 0}$ is set up to
nonperturbatively determine the hadronic bound states,
and $H_{\lambda I}$ should be treated perturbatively.}  This
separation is a basic step to realize a weak-coupling
treatment of nonperturbative QCD \cite{Wilson94}.
Thus, besides the free quark and gluon Hamiltonian
with constituent masses, $H_{\lambda 0}$
also contains the instantaneous interaction and all
the second order contributions generated by integrating
over all the modes above the low energy cutoff $\lambda$,
namely, eqs.(\ref{sfl}) and (\ref{vvl}).  Explicitly,
\begin{equation}
	H_{\lambda 0 ij} = \theta({\lambda^2 / K^+} - |\Delta
		K_{ij}^-| ) \Big\{ H_{Q\overline{Q}free}
		+ V_{Q\overline{Q}I} \Big\} , \label{QQeh}
\end{equation}
where
\begin{eqnarray}
	H_{Q\overline{Q}free} &&= [2(2\pi)^3]^2 \delta^3
		(\bar{k}_1-\bar{k}_3) \delta^3(\bar{k}_2-\bar{k}_4)
	 	\delta_{\lambda_1 \lambda_3} \delta_{\lambda_2
		\lambda_4} \nonumber \\
	& & ~~~~~~~~~~~~~~~ \times \Bigg\{ k_1^- + k_2^- +{\vec{k}_1^2
		\over 2m_Q}+{\vec{k}_2^2
		\over 2m_{\overline{Q}}} - 2{g^2 \over 4\pi^2}
		C_f {\lambda^2 \over K^+}\ln \epsilon \Bigg\} ,
			\label{QQfe} \\
	V_{Q\overline{Q}} (y-y', && \kappa_\bot
		-\kappa'_\bot) =  2(2\pi)^3 \delta^3(\bar{k}_1
		+ \bar{k}_2 - \bar{k}_3 - \bar{k}_4) \delta_{\lambda_1
		\lambda_3} \delta_{\lambda_2 \lambda_4}
		{-4g^2 (T^a)(T^a) \over (K^+)^2}
		\Bigg\{{1 \over (y-y')^2} \nonumber \\
	&& ~~~~ + {1 \over (y-y')^2} {\kappa_{\bot} - \kappa_{\bot}')^2
		\over (\kappa_{\bot} - \kappa_{\bot}')^2 + (y-y')^2
		\overline{\Lambda}^2} \theta(A(y-y', \kappa_\bot -
		\kappa'_\bot, \overline{\Lambda}) - \lambda^2) \Bigg\}
		\nonumber \\
	&& =  2(2\pi)^3 \delta^3(\bar{k}_1 + \bar{k}_2 - \bar{k}_3 -
		\bar{k}_4) \delta_{\lambda_1 \lambda_3} \delta_{\lambda_2
		\lambda_4}  \nonumber \\
	&& ~~~~~ \times {-4g^2 (T^a)(T^a) \over (K^+)^2} \Bigg\{{1
		\over (y-y')^2} \Big(1 - \theta( A(y-y', \kappa_\bot
		- \kappa'_\bot, \overline{\Lambda}) - \lambda^2) \Big)
		\nonumber \\ &&
	 ~~~~~~~~~ + {\overline{\Lambda}^2 \over (\kappa_{\bot} -
		\kappa_{\bot}')^2 + (y-y')^2 \overline{\Lambda}^2}
		\theta(A(y-y', \kappa_\bot - \kappa'_\bot,
		\overline{\Lambda}) - \lambda^2)
		\Bigg\}  \label{QQvv}.
\end{eqnarray}
In (\ref{QQfe}), $k_i^-$ is given by (\ref{k-}),
$\vec{k}_i^2 = {2 \over v^+} (k_\bot^2 - {1\over v^+}
2v_\bot \cdot k_\bot k^+ + {v^-\over v^+} k^{+2})$.
In the above results, we have already let UV cutoff
parameter $\Lambda \rightarrow \infty$ and the associated
divergence has been put in the mass correction. The kinetic
energy (\ref{hke}) now is included in the above nonperturbative
part of the effective Hamiltonian.

The free energy part $H_{Q\overline{Q} free}$ of eq.(\ref{QQfe})
has also included the self-energy correction which is the
instantaneous interaction contribution, a normal ordering
term of the instantaneous interaction in (\ref{lll}),
plus the one-loop contribution (\ref{sfl}). The result is
\begin{eqnarray}
	\Sigma = 4g^2 C_f \int && {dy_1 d^2 \kappa_{1\bot}
		\over 2(2\pi)^3} \Bigg\{ {\theta(y-y_1) \over
		(y-y_1)^2} \nonumber \\
	  && ~~~~~~~ - { \theta(y-y_1)\over (y-y_1)^2 }
		{(\kappa_{\bot}-\kappa_{1\bot})^2 \over
		(\kappa_{\bot}-\kappa_{1\bot})^2 + (y-y_1)^2
		\overline{\Lambda}^2}F(y-y_1, \kappa_\bot -
		\kappa_{1\bot}, \overline{\Lambda}) \Bigg\}   \nonumber \\
	  && \stackrel{\Lambda \rightarrow \infty}{=}
		  4g^2 C_f \int {dy_1 d^2 \kappa_{1\bot} \over
		2(2\pi)^3} {\theta(y-y_1)\over (y-y_1)^2} \theta(\lambda^2 -A
		(y-y_1, \kappa_\bot - \kappa_{1\bot}, \overline{\Lambda}))
		+ \delta m_Q^2  \nonumber \\
	  && = -{g^2 \over 4\pi^2}\lambda^2 C_f \ln \epsilon
		+ \delta m_Q^2 ,	\label{cqse}
\end{eqnarray}
where the mass correction $\delta m_Q^2 = {g^2 \over 4\pi^2} C_f
\overline{\Lambda}^2\ln {\Lambda^2 \over \lambda^2}$ which
has been renormalized away in eq.(\ref{QQfe}). By removing
away this mass correction, we should assign the corresponding
constituent quark mass in $H_{\lambda 0}$ being $\lambda$-dependent.
But, the heavy quark mass is much larger than
the low energy scale.  Its dependence on $\lambda$
should be very weak and could be neglected.

The $Q\overline{Q}$ interaction $V_{Q\overline{Q}}$ of
eq.(\ref{QQvv}) contains the one gluon exchange interaction
 eq.(\ref{vvl}) plus the instantaneous interaction [the
last term in (\ref{lll})].  It clearly shows that without the
 low energy cutoff ($\lambda=0$), the instantaneous interaction
is completely cancelled
by the same contribution from the one transverse gluon exchange
and the  remaining one gluon exchange interaction is a
Coulomb interaction, like in QED.  Now, with the low energy
cutoff, the one gluon exchange contribution only contains
these gluons with the energy greater than the cutoff $\lambda$.
Thus, the resulting $Q\overline{Q}$ interaction has two terms:
The first term is the result of the noncancellation between
the instantaneous interaction and one transverse gluon
exchange interaction, which corresponds to a confining potential.
The second term, the rest of one transverse gluon exchange
interaction, is the Coulomb interaction on the light-front.
The detailed confinement mechanism on the light-front will be
discussed in section V. With the kinetic energy incorporated,
we see that the above effective
QCD Hamiltonian which will be used to determine the heavy
quarkonium bound states only has the spin symmetry but no
flavour symmetry.

Before ending this section, we may compare the present
formulation for heavy quarkonia with the nonrelativistic
QCD formulation developed by Lepage et al. \cite{Lepage95}.

In the nonrelativistic QCD formulation,
heavy quarkonia are described by an effective field theory of
QCD in the nonrelativistic limit plus a systematic computations
of the relativistic corrections (in terms of momentum scales
($Mv)^2$ and ($Mv^2)^2$) and QCD short distance corrections
(in terms of the scale $M^2$). The nonperturbative QCD scale
$\Lambda_{QCD}^2$ is implicatively included in this formulation.
Heavy quarkonium annihilation and production processes can
then be factorized with respect of the above different scales.
The dominant contributions in quarkonium processes,
namely, the nonperturbative QCD dynamics of quarkonium bound
states may be computed in lattice simulations.

Our formulation is based on the factorization scale $m_Q$
which naturally separates the QCD short distance
and long distance dynamics. The long distance dynamics is
described by the residual momentum which is now controlled
by $\Lambda^2_{QCD}/\lambda^2$ via the effective Hamiltonian
$H_{\lambda 0}$. The resulting effective Hamiltonian derived
from QCD by the similarity renormalization group
approach contains explicitly the confining and Coulomb
interactions which have encompassed the necessary long
distance effects for heavy quarkonia. The quarkonium bound
states can then be directly solved in the corresponding
light-front bound state equation (as we shall see later).
The short distance dynamics can be systematically computed
in the ordinary perturbation theory. These include the
QCD radiative corrections (controlled by $\ln(m_Q/\Lambda)$,
where $\Lambda$ is an UV cutoff), and the $1/m_Q$ corrections
(controlled by $\overline{\Lambda}/m_Q$). Our formulation is
fully relativistic. It is a more complete QCD formulation
for heavy quarkonia in comparison to nonrelativistic QCD
\cite{Lepage95}.  It allows to directly compute the nonperturbative
QCD dynamics without the help of lattice simulation. Moreover,
it is also straightforward to extend this formulation to the heavy
hadron system which contains a single heavy quark, as we shall
see the next.

\subsection{Low-energy effective Hamiltonian for
	heavy-light quark systems}

The heavy-light quark system (heavy hadrons with one
heavy quark) is  one of the most interesting
topics in the current study of heavy hadron physics.
We now apply the similarity renormalization group
approach to such system.

The bare cutoff Hamiltonian we begin with for heavy-light
quark system is the combination of the heavy
quark effective Hamiltonian (\ref{hqbeh}) and the
full Hamiltonian for the light quark (\ref{qcdh}).
Due to the HQET, we may also introduce
the residual center mass momentum for heavy-light systems,
\begin{equation}
	K^+ = \overline{\Lambda} v^+ = p^+_1 + k^+_1 = p^+_2 + k^+_2,  ~~~
	~~~ K_\bot = \overline{\Lambda} v_\bot = p_{1\bot} + k_{1\bot}
		 = p_{2\bot} + k_{2\bot},
\end{equation}
where $\overline{\Lambda} = M_H-m_Q$, $p_1$ and $p_2$ are
the light antiquark momenta in the initial and final states
respectively, and $k_1$ and $k_2$ the residual
momenta of the heavy quarks.  The initial and final states
in $Q\overline{q}$ sector are denoted by $| i \rangle
= b^\dagger_v(k_1,\lambda_1) d^\dagger (p_1,\lambda_1')
|0 \rangle$ and  $| j \rangle = b^\dagger_v(k_2,\lambda_2)
d^\dagger (p_2,\lambda_2')|0 \rangle$, respectively.
The residual longitudinal
momentum fractions and the residual relative transverse
momenta are defined in the similar way as in quarkonium
system,
\begin{eqnarray}
	&& y = p^+_1 / K^+ ~,~~~~~ \kappa_{\bot} =
		p_{1\bot} - y K_{\bot} , \nonumber \\
	&& y' = p^+_2 / K^+ ~,~~~~~ \kappa'_{\bot} =
		p_{2\bot} - y' K_{\bot},
\end{eqnarray}
but the range of the longitudinal momentum fractions $y$
and $y'$ is different:
\begin{equation}
 0 < y = {M_H \over \overline{\Lambda}} {p_1^+ \over P^+} < \infty~~,
	~~ 0 < y' = {M_H \over \overline{\Lambda}} {p_2^+
		\over P^+} < \infty .	\label{Qqlm}
\end{equation}

Following the general procedure described in the previous subsection,
it is easy to find that the nonperturbative part of the
low-energy effective Hamiltonian for heavy-light quark systems,
which is given by
\begin{equation}
	H_{\lambda 0 ij} = \theta({\lambda^2 \over K^+} -|\Delta
		K_{ij}^-| ) \Big\{ H_{Q\overline{q}free}
		+ V_{Q\overline{q}I} \Big\} , \label{Qqeh}
\end{equation}
where
\begin{eqnarray}
	H_{Q\overline{q}free} &&= [2(2\pi)^3]^2 \delta^3
		(\bar{k}_1-\bar{k}_2) \delta^3(\bar{p}_1-\bar{p}_2)
	 	\delta_{\lambda_1 \lambda_3} \delta_{\lambda_2
		\lambda_4}	 \nonumber \\
	& & ~~~~~\times \Bigg\{ {p^2_{\bot} + m_q^2 \over p^+} +
		{1\over K^+} \Big(2 K_{\bot} \cdot k_{\bot} - K^-k^+
		\Big) - {g^2 \over 2\pi^2 } C_f {\lambda^2 \over K^+}
		\ln \epsilon \Bigg\} , \label{Qqeh1} \\
	V_{Q\overline{q}}(y-y', && \kappa_\bot - \kappa'_\bot)
		=  2(2\pi)^3 \delta^3(\bar{k}_1 + \bar{p}_1 - \bar{k}_2 -
		\bar{p}_2) \delta_{\lambda_1 \lambda_3} \delta_{\lambda_2
		\lambda_4}	\nonumber \\
	& & ~~~~~ \times  (- 2g^2) (T^a)(T^a) \Bigg\{ 2\Bigg({1\over
		q^+}\Bigg)^2 + { 1 \over q^+} \Bigg({ q^{i'}_{\bot}
		\over q^+} - {v^{i'}\over v^+}\Bigg) \Bigg(2{q^{i'}_\bot
		\over q^+} - {p_{1\bot}^{i'} \over p_1^+} -
		{p_{2\bot}^{i'} \over p_2^+} \Bigg) \nonumber \\
	& & ~~~~~~~ \times \theta({\Lambda^2 \over K^+} - |p_1^- - p_2^-
		- q^-|) \theta({\Lambda^2 \over K^+} - |k_2^- - k_1^-
		- q^-|) \nonumber \\
	& & ~~~~~~~ \times \Bigg[{\theta(|p_1^- - p_2^- - q^-| -
		{\lambda^2 \over K^+})\theta(|p_1^- - p_2^- - q^-| -
		|k_2^- - k_1^- - q^-| \over p_1^-
		- p_2^- - q^-} \nonumber \\
	& & ~~~~~~~~~~+ {\theta(|k_2^- - k_1^- - q^-|-{\lambda^2 \over K^+})
		\theta(|k_2^- - k_1^- - q^-| - |p_1^- - p_2^- - q^-|
		\over k_2^- - k_1^- - q^-} \Bigg] \Bigg\} \nonumber \\
	&&{\stackrel{\Lambda \rightarrow \infty}{=}}  2(2\pi)^3
		\delta^3(\bar{k}_1 + \bar{p}_1 - \bar{k}_2 - \bar{p}_2)
		\delta_{\lambda_1 \lambda_3} \delta_{\lambda_2 \lambda_4}
		{- 2g^2 (T^a)(T^a) \over (K^+)^2}
		\nonumber \\
	&&~~~~~~~~ \times  \Bigg\{ {2\over (y-y')^2}
		- \Bigg[ 2{(\kappa_\bot-\kappa'_\bot)^2 \over
		(y - y')^2} - {\kappa^2_\bot -\kappa_\bot \cdot
		\kappa'_\bot \over y(y-y') } - {\kappa_\bot \cdot
		\kappa'_\bot - (\kappa')^2_\bot \over y'(y-y')}
		\Bigg] \nonumber \\
	& & ~~~~~~~~~~~~~~~~~~~~~~~~ \times \Bigg[ {\theta(B - \lambda^2)
		\theta(B - A) \over (\kappa_\bot-\kappa'_\bot)^2
		-{y-y'\over y}(\kappa_\bot^2 + m_q^2) -
		{y-y' \over y'} ((\kappa')_\bot^2 + m_q^2)}
		\nonumber \\
	& &~~~~~~~~~~~~~~~~~~~~~~~~~~~~~~~~~~~~~~~~~
		+ {\theta(A - \lambda^2) \theta(A - B) \over
		(\kappa_\bot-\kappa'_\bot)^2 + (y-y')^2 \overline{
		\Lambda}^2} \Bigg] \Bigg\}, \label{Qqvv}
\end{eqnarray}
with
\begin{eqnarray}
	&& B \equiv \Bigg|{(\kappa_\bot-\kappa'_\bot)^2 \over
		y-y'} - {\kappa_\bot^2 + m_q^2 \over y}
		- {(\kappa')_\bot^2 + m_q^2 \over y'} \Bigg| , \\
	&& A \equiv (\kappa_\bot -\kappa'_\bot)^2/|y-y'|
		+ |y-y'| \overline{\Lambda}^2 .
\end{eqnarray}
Here we do not include the heavy quark kinetic energy into
$H_{\lambda 0}$ since the dominant kinetic energy is given
by the constituent light quark. The heavy quark kinetic energy
can be treated as a perturbative correction to $H_{\lambda 0}$.
The heavy quark free energy has been written in (\ref{Qqeh1})
by
\begin{equation}
	k^- = {1 \over v^+} \Big(2v_\bot \cdot k_\bot - v^- k^+\Big)
		= {1 \over K^+} \Big(2K_\bot \cdot k_\bot - K^- k^+\Big).
\end{equation}
The low energy heavy-light quark effective Hamiltonian
is $m_Q$-independent. It obviously has the spin and flavour
 symmetry, namely, the heavy quark symmetry.  Comparing
to the quarkonium systems, the $Q\overline{q}$ interactions
are much more complicated. But it is not difficult to see that
the above $H_\lambda$ contains a confining potential.

Finally, it is also straightforward to extend the above
derivation to light-light quark systems.  The result is
just eq.(\ref{eh}) but in terms of the relative momenta:
\begin{eqnarray}
	x = p_1^+ /P^+~~, ~~~~ \kappa_\bot = p_{1\bot}
		- x P_\bot , \nonumber \\
	x'= p_3^+ /P^+ ~~, ~~~~ \kappa'_\bot = p_{1\bot}
		- x' P_\bot .
\end{eqnarray}
Here there is no residual center mass momentum for
light-light system. The hadron momentum $P^\mu$ is already
of order a low energy scale.
We shall not intend to discuss the light-light
systems in details in this paper.
As we have pointed out in the Introduction,
for the light-light quark systems, besides the confinement,
chiral symmetry breaking also plays an essential role in
the low energy hadronic dynamics.  We shall remain the
light-light quark systems for further investigation when
we attempt to address the problem of chiral symmetry in
light-front QCD \cite{Zhang}.

In conclusion, we have obtained in this section the
renormalized low energy effective QCD Hamiltonian for
heavy-heavy and heavy-light quark systems, and
extracted the nonperturbative part $H_{\lambda 0}$,
eqs.(\ref{QQeh}) and (\ref{Qqeh}), in the weak-coupling
treatment scheme.  We are now ready to solve heavy
hadrons on the light-front.

\section{Light-front heavy hadron bound state equations}

In this section, based on the low-energy heavy quark effective
Hamiltonian derived in the previous section, we shall
construct light-front bound state equations in the weak-coupling
treatment scheme.

\subsection{General structure of light-front bound state
equations in the weak-coupling treatment scheme}

In general, a hadronic bound state on the light-front can
be expanded in the Fock space composed of states with definite
number of particles \cite{Brodsky81,Zhang94a}. Formally, it
can be expressed as follows
\begin{equation}
        | \Psi(P^+, P_{\bot},\lambda_s) \rangle = \sum_{n,\lambda_i}
                \int \Big( \prod_i [d^3 \bar{p}] \Big)
                2 (2\pi)^3 \delta^3(\bar{P}-\sum_i \bar{p}_i)
                | n, \bar{p},\lambda_i \rangle
                \Phi_{n} (x_i,\kappa_{\bot i},\lambda_i), \label{lfwf}
\end{equation}
where $P^+, P_{\bot}$ are its total longitudinal and transverse
momenta respectively and $\lambda_s$ its total helicity, $|n,
\bar{p}, \lambda_i \rangle$ is a Fock state consisting of $n$
constituents, each of which carries momentum $\bar{p}_i$ and
helicity $\lambda_i$ ($\sum_i \lambda_i = \lambda_s$);
$\Phi(x_i,\kappa_{\bot i},\lambda_i)$ the corresponding amplitude
which depends on the helicities $\lambda_i$, the longitudinal
momentum fractions $x_i$, and the relative transverse momenta
$\kappa_{\bot i}$:
\begin{equation}
	x_i = { p_i^+ \over P^+}~~,
		~~~ \kappa_{i\bot} = p_{i\bot} - x_i P_{\bot}.
\end{equation}

The eigenstate equation that the wave functions obey on the light-front
is obtained from the operator Einstein equation $P^2 =P^+P^- - P_{\bot}^2
= M^2$:
\begin{equation}
        H_{LF} | P^+, P_{\bot},\lambda_s \rangle  = { P_{\bot}^2
                + M^2 \over P^+ } | P^+, P_{\bot},
                \lambda_s \rangle ,
\end{equation}
where $H_{LF}={P}^-$ is the light-front Hamiltonian. Explicitly, for
a meson wave function, the corresponding light-front bound state
equation is:
\begin{equation}
        \Big(M^2 - \sum_i { \kappa_{i\bot}^2 + m_i^2 \over x_i} \Big)
                \left[\begin{array}{c} \Phi_{q\overline{q}} \\
                \Phi_{q\overline{q}g} \\ \vdots \end{array} \right]
                  = \left[ \begin{array}{ccc} \langle q \overline{q}
                | H_{int} | q \overline{q} \rangle & \langle q
		\overline{q} | H_{int} | q \overline{q} g \rangle
		& \cdots \\ \langle q \overline{q} g
                | H_{int} | q \overline{q} \rangle & \cdots & ~~  \\ \vdots &
                \ddots & ~~ \end{array} \right] \left[\begin{array}{c}
                \Phi_{q\overline{q}} \\ \Phi_{q\overline{q}g} \\
		\vdots \end{array} \right], \label{lfbe}
\end{equation}
where $H_{int}$ is the interaction part of $H_{LF}$.

Obviously, solving eq.(\ref{lfbe}) from QCD with the entire
Fock space is impossible.  A basic motivation of introducing
the weak-coupling treatment scheme
is to simplify the complexities in solving the above equation.
The Hamiltonian $H_\lambda$ derived in the previous sections has
already decoupled the high and low energy states.
Here we only consider the low energy states (hadronic bound
states).  Due to the kinematic feature of boost
symmetry on the light-front, we can
assign a relative small longitudinal light-front momentum to
the bound states.  On the other hand, the light-front infrared
divergences force us to introduce a small cutoff on the
longitudinal light-front momentum to each individual constituent.
Thus, the hadronic bound states can only consist of the Fock
space sectors with a few particles. This is a kinematic
truncation on eqs.(\ref{lfwf}) and (\ref{lfbe}). Furthermore,
the most important point in the weak-coupling treatment scheme is
the reseparation of the Hamiltonian $H_\lambda
=H_{\lambda 0} + H_{\lambda I}$. As we mentioned before,
$H_{\lambda 0}$ which conserves particle number devotes
to a nonperturbative evaluation to the bound states
through eq.(\ref{lfbe}). And $H_{\lambda I}$ which
describes the particle emissions and reabsoptions is hopefully
a perturbative term in the weak-coupling treatment so that
we may not consider its contribution to eq.(\ref{lfbe}).
Then, eq.(\ref{lfbe}) becomes diagonal in Fock
space with respect to the different particle number sectors,
\begin{equation}
        \Big(M^2 - \sum_i { \kappa_{i\bot}^2 + m_i^2 \over x_i} \Big)
                \left[\begin{array}{c} \Phi_{q\overline{q}} \\
                \Phi_{q\overline{q}g} \\ \vdots \end{array} \right]
                  = \left[ \begin{array}{ccc} \langle q \overline{q}
                | H_{\lambda 0} | q \overline{q} \rangle & 0
		& ~~~~~ 0 \\ 0 & \langle q \overline{q} g
                | H_{\lambda 0} | q \overline{q} g\rangle & ~~~~~ 0  \\ 0 &
                0 & \ddots \end{array} \right] \left[\begin{array}{c}
                \Phi_{q\overline{q}} \\ \Phi_{q\overline{q}g} \\
		\vdots \end{array} \right]. \label{lfbe1}
\end{equation}
Now we see that the bound state equation is manable.

The second important step in the weak-coupling
treatment to the low energy QCD is the use of a constituent
picture. The success of the constituent quark model
suggests that we may only consider the valence
quark Fock space in determining the hadronic bound
states from $H_{\lambda 0}$.  In this picture,
quarks and gluons must have constituent masses.
This constituent picture can naturally be realized on
the light-front \cite{Wilson94}. However, an essential
difference from the phenomenological constituent
quark model description is that the constituent masses
introduced here are $\lambda$ dependent.  This cutoff
dependence of constituent masses (as well as the effective
coupling constant) is determined by solving the bound
states equation and fitting the physical quantities with
experimental data. This is indeed a renormalization
condition in nonperturbative QCD.  Note that unlike
the usual renormalization scheme in QED, quarks and gluons
in QCD are not physically observable particles so that
we can only determine their renormalized masses and
coupling constant in hadronic (composite particles) sectors.
Once the constituent picture is introduced,
we can truncate the general expression of the light-front
bound states to only including the valence quark Fock
space.  The higher Fock space contributions can be
calculated as a perturbative correction through $H_{\lambda I}$.
Thus, for mesons,
eq.(\ref{lfwf}) is reduced to the following simple form:
\begin{eqnarray}
        |\Psi(P^+,P_{\bot},\lambda_s) \rangle = \sum_{\lambda_1 \lambda_2}
        \int [d^3\bar{p}_1] [d^3\bar{p}_2] && 2 (2\pi)^3
		\delta^3(\bar{P}-\bar{p}_1-\bar{p}_2) \nonumber \\
	&& \times \Phi_{q\overline{q}}(x,\kappa_{\bot}
		\lambda_1, \lambda_2)
		| q(p_1,\lambda_2) \overline{q}(p_2,\lambda_2)
		\rangle,  \label{lfwcwf}
\end{eqnarray}
where $| q(p_1,\lambda_1) \overline{q}(p_2,\lambda_2) \rangle
=b^\dagger(p_1,\lambda_1) d^\dagger(p_2,\lambda_2) |0\rangle$,
and $b^\dagger, d^\dagger$ should be regarded as the creation
operator of the constituent quark and antiquark respectively.
Consequently, the constituent quark and gluon masses $m_i$
and  coupling constant $g$ in the effective
Hamiltonian $H_{\lambda 0}$ become explicit functions
of the low energy cutoff $\lambda$.

It is worth pointing out that spin is always a troublesome issue
in the light-front approach. The meson light-front bound state
we have constructed is labelled by helicity rather than spin.
However in practice low-energy hadronic states with definite
spins are needed. This discrepancy is usually remedied by introducing
the so-called Melosh rotation \cite{Melosh}, which transforms
a single particle state from the light-front helicity basis
to the ordinary spin basis,
\begin{equation}
        R(x_i,k_{\bot},m_i) =  { m_i + x_i M_0 - i \sigma \cdot
                ({\bf n} \times \kappa_{\bot} ) \over \sqrt{ (m_i +
                x_i M_0)^2 + \kappa_{\bot}^2}} , \label{4.10}
\end{equation}
where ${\bf n} = (0, 0, 1)$, and
\begin{equation}
        M_0^2 = {\kappa_{\bot}^2 + m_1^2 \over x} + {\kappa_{\bot}^2
		+ m_2^2 \over 1-x} .
\end{equation}
With Melosh transformation incorporated, the light-front
meson bound state with a definite spin can be expressed in
the weak-coupling treatment scheme as follows
\begin{eqnarray}
        | \Psi (P^+, P_{\bot},J,J_z) \rangle = \sum_{\lambda_1\lambda_2}
        \int [d^3\bar{p}_1][d^3\bar{p}_2] && 2(2\pi)^3
		\delta^3(\bar{P}-\bar{p}_1- \bar{p}_2) \nonumber \\
	&& \times \Phi^{JJ_z}_{q\overline{q}}(x,\kappa_{\bot},
		\lambda_1, \lambda_2)| q(p_1,\lambda_1)
		\overline{q}(p_2,\lambda_2)\rangle,
                                                  \label{lfwcwf1}
\end{eqnarray}
where
\begin{eqnarray}
	&& \Phi^{JJ_z}_{q\overline{q}}(x,\kappa_{\bot},
		\lambda_1, \lambda_2) = \phi_{q\overline{q}}
		(x,\kappa_{\bot}) R^{JJ_z}_{\lambda_1 \lambda_2}
                (x,\kappa_{\bot}), \\
        && R_{\lambda_1 \lambda_2}^{JJ_z} (x,\kappa_{\bot}) = \sum_{s_1s_2}
                 \langle \lambda_1 | R^{\dagger} (x, \kappa_{\bot}, m_1)
                 | s_1 \rangle \langle \lambda_2 | R^{\dagger} (1-x,
                 - \kappa_{\bot}, m_2) | s_2 \rangle \langle {1\over 2}
                 s_1 {1\over 2}s_2 | JJ_z \rangle,   \label{mttb}
\end{eqnarray}
and $\langle {1\over 2}s_1 {1\over 2}s_2 | JJ_z \rangle$ is the
Clebsch-Gordon coefficient.
The normalization condition for the state $|\Psi(v,J,J_z)\rangle$
is taken to be
\begin{equation}
        \langle \Psi({P'}^+, P'_\bot ,J',J'_z) | \Psi(P^+, P_\bot,J,J_z)
		\rangle = 2(2\pi)^3 P^+ \delta^3(\bar{P}'-\bar{P})
                \delta_{J'J}\delta_{J'_z J_z}, \label{4.17}
\end{equation}
which leads to
\begin{equation}
       \int {dx d^2 \kappa_{\bot} \over 2 (2\pi)^3}
                |\phi_{q\overline{q}}(x,\kappa_{\bot})|^2 = 1. \label{4.18}
\end{equation}
After the above consideration of the spin property on the
light-front, eq.(\ref{lfbe}), for mesons, becomes
a light-front Bethe-Salpeter equation:
\begin{eqnarray}
        \Big( M^2 - M_0^2 \Big) \Phi^{JJ_z}_{q \overline{q}}
		(x,\kappa_{\bot}, \lambda_1, \lambda_2)
	=&& \Bigg(-{g_\lambda^2 \over 2\pi^2}  \lambda^2 C_f \ln \epsilon
		\Bigg) \Phi^{JJ_z}_{q \overline{q}} (x,\kappa_{\bot},
		\lambda_1, \lambda_2) \nonumber \\
	&& + \sum_{\lambda'_1 \lambda'_2}\int {dx' d^2
		\kappa'_{\bot} \over 2 (2\pi)^3}
           	V_{eff} (x,\kappa_{\bot},\lambda_1, \lambda_2;
		x', \kappa'_{\bot}, \lambda'_1, \lambda'_2) \nonumber \\
	&& ~~~~~~~~~~~~~ \times \Phi^{JJ_z}_{q \overline{q}}
		(x', \kappa'_{\bot}, \lambda'_1, \lambda'_2), \label{lfbse}
\end{eqnarray}
where $V_{eff}$ is the effective $q\overline{q}$ interactions in
eq.(\ref{eh}).

Melosh transformation is exact only for free theory.
With interactions incorporated, the use of Melosh transformation
is only an approximation.  This approximation may be reasonably
good for the lowest spin bound states, such as the lowest-lying
scalar and vector mesons, since the  wavefunction (the valence
quark amplitude in the light-front bound states) for them
is a scale function to the rotational transformation.
In other words, the ``orbit'' angular momentum which is dynamically
dependent in the light-front formulation may not contribute
to the total spin of these lowest spin hadrons. Moreover,
in this paper, we focus on heavy quark systems.
As we shall see next, due to the heavy quark spin symmetry,
Melosh transformation results in the exact spin structure
with fixed parity for scalar and vector heavy mesons.

\subsection{Bound state equation for heavy quarkonia}

In this section, we shall explicitly consider the heavy
quarkonium states. First of all, for quarkonia,
the wave function (\ref{lfwcwf1})
can be further simplified, especially for its spin structure
due to the spin symmetry in HQET.  Within the
framework of light-front HQET, eq.(\ref{lfwcwf1}) in the
heavy quark limit is reduced to:
\begin{eqnarray}
        | \Psi (K, J, J_z) \rangle = \sum_{\lambda_1 \lambda_2}
		R_{\lambda_1 \lambda_2}^{JJ_z} \int [d^3\bar{k}_1]
		[d^3\bar{k}_2] && 2(2\pi)^3 \delta^3(\bar{K} -
		\bar{k}_1-\bar{k}_2) \nonumber \\
		&& \times \phi_{Q\overline{Q}}
         	(y,\kappa_{\bot}) | b^\dagger_v(k_1,\lambda_1)
		d^\dagger_v(k_2, \lambda_2) \rangle .   \label{QQwf}
\end{eqnarray}
Here the wavefunction $\phi_{Q\overline{Q}}(y,\kappa_{\bot})$
may be mass dependent due to the kinetic energy in $H_{\lambda 0}$
[see (\ref{QQfe})]. The Melosh transformation matrix element
(\ref{mttb}) in quarkonium states becomes a pure kinematic factor,
\begin{equation}
        R_{\lambda_1 \lambda_2}^{00} = { v^+ \over 2\sqrt{2}}~
		\overline{u}(v, \lambda_1) \gamma^5 v (v,\lambda_2)
		\label{4.14}
\end{equation}
for a pseudoscalar meson, and
\begin{equation}
        R_{\lambda_1 \lambda_2}^{1J_z} = - { v^+ \over 2\sqrt{2}}~
		\overline{u}(v, \lambda_1) \! \not{\! {\epsilon}}(J_z)
		v(v,\lambda_2)  \label{4.19}
\end{equation}
for vector mesons.  The light-front spinors for heavy quarks are
given by
\begin{eqnarray}
	&& u(v,\lambda) = \Big(1 + {\alpha \cdot v_{\bot} + \beta
		\over v^+ } \Big) w_{\lambda} = \left(
		\begin{array}{c} 1 \\ {1\over v^+} (\tilde{\sigma}
		\cdot v_\bot +i) \end{array} \right) \chi_{\lambda},
			\nonumber \\
	&& v(v,\lambda) = \Big(1 + {\alpha \cdot v_{\bot} - \beta
		\over v^+ } \Big) w_{-\lambda} = \left( \begin{array}{c}
		1 \\ {1\over v^+} (\tilde{\sigma} \cdot v_\bot -i)
		\end{array} \right) \chi_{-\lambda},
\end{eqnarray}
so that
\begin{eqnarray}
	&& \overline{u}(v,\lambda)u(v,\lambda') = {2 \over v^+}
		\delta_{\lambda \lambda'}~~, ~~~
	 \sum_{\lambda}u(v,\lambda)\overline{u}(v,\lambda) =
		{1+\not{\! v} \over v^+} , \\
	&& \overline{v}(v,\lambda)v(v,\lambda') = -{2 \over v^+}
		\delta_{\lambda \lambda'}~~, ~~~
	 \sum_{\lambda}v(v,\lambda)\overline{v}(v,\lambda) = -
		{1-\not{\! v} \over v^+} ,
\end{eqnarray}
and the polarization vector is defined by
\begin{equation}
        {\epsilon}^{\mu} (\pm1) = \Big( {2 \over v^+}
         \epsilon_{\bot} \cdot v_{\bot}, 0, \epsilon_{\bot} \Big)~,~~
        {\epsilon}^\mu (0) = - \Big( {v^2_{\bot} -1 \over v^+}, v^+,
                v_{\bot} \Big) ~, ~~ \epsilon_\bot (\pm 1) =
		\mp {1 \over \sqrt{2}} (1 \pm i).     \label{polv}
\end{equation}
Thus we have constructed the light-front heavy quarkonium bound
states in the heavy mass limit, which have definite
spin and parity.  The corresponding spin tensor
structures are given by eqs.(\ref{4.14}) and ({\ref{4.19}).

Note that the heavy quarkonium states in heavy mass limit
are labelled by the residual center mass momentum $K^\mu$.
We may normalize eq.(\ref{QQwf}) as follows:
\begin{equation}
        \langle \Psi(K',J',J'_z) | \Psi(K,J,J_z)
		\rangle = 2(2\pi)^3 K^+ \delta^3(\bar{K}-\bar{K}')
                \delta_{J'J}\delta_{J'_z J_z}, \label{nQQbs}
\end{equation}
which leads to
\begin{equation}
       \int {dy d^2\kappa_{\bot} \over 2 (2\pi)^3} |\phi_{q\overline{q}}
		(y,\kappa_{\bot})|^2 = 1. \label{nQQwf}
\end{equation}

With the quarkonium states given above, it is easy to derive
the corresponding
bound state equation.  In the weak-coupling scheme, $H_{LF} = H_{
\lambda 0}$ where $H_{\lambda 0}$ is given by (\ref{QQeh}). Thus, the
quarkonium bound state equation in light-front HQET is given by
\begin{equation}	\label{QQb}
	(K^- - H_{\lambda 0}) |\Psi(P,J,J_z) \rangle = 0 .
\end{equation}
The free energy part of the quarkonia states is extremely simple,
\begin{eqnarray}
	 K^- - k_1^- - k_2^- = {1 \over K^+}
		2 \overline{\Lambda}^2 \equiv {1\over K^+}
		(\overline{\Lambda}^2 - \overline{M}_0^2) ,
\end{eqnarray}
where $\overline{M}_0^2 = - \overline{\Lambda}^2$ is a residual
invariant mass (=the invariant mass $M_0^2$ subtracted by the mass
dependent terms,  here $\overline{M}^2$ is just a notation rather
than a real square of a quantity). It follows that
eq.(\ref{QQb}) can be expressed explicitly by
\begin{eqnarray}
     \Bigg\{2\overline{\Lambda}^2 - {\overline{\Lambda}\over m_Q}
	\Big[2\kappa_\bot^2 + && \overline{\Lambda}^2(2y^2-2y +1)
	\Big] \Bigg\} \phi_{Q\overline{Q}}(y,k_{\bot})
	= \Bigg(-{g_\lambda^2\over 2 \pi^2} \lambda^2 C_f \ln\epsilon\Bigg)
		~\phi_{Q\overline{Q}}(y,k_{\bot}) \nonumber \\
	&& ~~ -4g_\lambda^2 (T^a)(T^a) \int {dy' d^2\kappa'_{\bot} \over
		2(2\pi)^3} \Bigg\{ {1 \over (y-y')^2} \theta(
		\lambda^2 - A) \nonumber \\ && ~~~~~~~~~~
	  +  {\overline{\Lambda}^2 \over (\kappa_{\bot} -
	  \kappa_{\bot}')^2 + (y-y')^2 \overline{\Lambda}^2} \theta
		(A - \lambda^2) \Bigg\} \phi_{Q\overline{Q}}
		(y',\kappa'_{\bot}).  \label{QQbse}
\end{eqnarray}
This is the bound state equation for heavy quarkonia in the
weak-coupling treatment of low energy QCD.

\subsection{Bound state equation for heavy-light quark systems}

In the previous subsection, we have derived explicitly the
bound state equation for quarkonia.  In the last few
years, the heavy-light quark systems have been extensively
explored theoretically and experimentally.  The
discovery of the heavy quark symmetry in the heavy mass limit
\cite{Isgur90} allows one to extract the
Kobayashi-Maskawa matrix element $|V_{cb}|$
without knowing the detailed structure of the heavy-light mesons.
However, to have a complete description
for various heavy hadron processes, one has to know a number of
heavy hadron matrix elements involved the heavy hadron bound
states. Currently, most of heavy hadron matrix elements
have only been calculated in various phenomenological
models, such as quark models, QCD sum rules etc. It is
necessary to find these heavy-light hadron bound
states from the fundamental QCD. In this
subsection, we shall derive from the low energy effective
QCD Hamiltonian the bound state equation obeyed
by the heavy mesons with one heavy quark.

For the heavy mesons with one heavy quark, due to the
heavy quark symmetry the general
form of the wavefunction in the weak-coupling scheme can
also be simplified.  Considering the heavy quark mass limit,
the wavefunction (\ref{lfwcwf1}) can be expressed as
\begin{eqnarray}
        | \Psi (K, J, J_z) \rangle = \sum_{\lambda_1 \lambda_2}
		\int [d^3\bar{k}] [d^3\bar{p}] && 2(2\pi)^3
		\delta^3(\bar{K} - \bar{k}-\bar{p}) \nonumber \\
		&& \times R_{\lambda_1 \lambda_2}^{JJ_z}(y,\kappa_\bot)
		\phi_{Q\overline{Q}}(y,\kappa_{\bot})
		| b^\dagger_v(k,\lambda_1), d^\dagger(p,
		\lambda_2) \rangle .    \label{Qqwf}
\end{eqnarray}
Here the Melosh transformation matrix element, eq.(\ref{mttb}),
is a little more complicated in comparing to the heavy
quarkonium state:
\begin{equation}
        R_{\lambda_1 \lambda_2}^{00} = {1\over 2}\sqrt{p^+ K^+ \over
		2(y\overline{\Lambda}^2 + {\kappa_\bot^2 +
		m_q^2(\lambda) \over y})}~
		\overline{u}(v, \lambda_1) \gamma^5 v (p,\lambda_2)
		\label{mtQq1}
\end{equation}
for a pseudoscalar meson, and
\begin{equation}
        R_{\lambda_1 \lambda_2}^{1J_z} = - {1\over 2}\sqrt{p^+ K^+ \over
		2(y\overline{\Lambda}^2 + {\kappa_\bot^2 + m_q^2(\lambda)
		\over y})}~
		\overline{u}(v, \lambda_1) \! \not{\! {\epsilon}}(J_z)
		v(p,\lambda_2)  \label{mtQq2}
\end{equation}
for vector mesons.  The light-front spinors for the light antiquark is
\begin{equation}
	v(p,\lambda) = \Big(1 + {\alpha \cdot p_{\bot} - \beta m_q(\lambda)
		\over p^+ } \Big) w_{-\lambda} = \left(\begin{array}{c}
		1 \\ {1\over p^+} (\tilde{\sigma} \cdot p_\bot
		-im_q(\lambda)) \end{array} \right) \chi_{-\lambda},
\end{equation}
and the polarization vector $\epsilon^\mu$ is still given by
eq.(\ref{polv}).

Eq.(\ref{Qqwf}) is a light-front heavy-light meson bound state
in the symmetry limit ($m_Q \rightarrow \infty$), which
has the definite spin and parity.  The corresponding bound state
equation then becomes
\begin{eqnarray}
     \Big(\overline{\Lambda}^2 + (1-y) \overline{\Lambda}^2
	- && {\kappa_\bot^2 + m^2_q(\lambda) \over y} \Big)
	\Phi^{JJ_z}_{Q\overline{q}}(y,k_{\bot},\lambda_1,
	\lambda_2) \nonumber \\
    && =  \Big(-{g_\lambda^2\over 2 \pi^2} \lambda^2 C_f \ln\epsilon\Big)
		~\Phi^{JJ_z}_{Q\overline{q}}(y,k_{\bot},
		\lambda_1, \lambda_2) \nonumber \\
	&& ~~~~ + (K^+)^2\int {dy' d^2\kappa'_{\bot} \over
		2(2\pi)^3} V_{Q\overline{q}}(y-y',\kappa_{\bot}
		- \kappa_{\bot}')\Phi^{JJ_z}_{Q\overline{q}}
		(y',\kappa'_{\bot}, \lambda_1, \lambda_2),
		\label{Qqbse}
\end{eqnarray}
where $V_{Q\overline{q}}$ is given by eq.(\ref{Qqvv}),
\begin{equation}
	\Phi^{JJ_z}_{Q\overline{q}}(y,\kappa_{\bot},\lambda_1,
		\lambda_2) = \phi_{Q\overline{q}}(y,\kappa_{\bot})
		R^{JJ_z}_{\lambda_1 \lambda_2}(y,\kappa_\bot) ,
\end{equation}
and $R^{JJ_z}_{\lambda_1 \lambda_2}$ is determined
by eq.(\ref{mtQq1}) or (\ref{mtQq2}).  Here
the light antiquark is a brown muck, a current
quark surround by infinite gluons and $q\overline{q}$
pairs that result a constituent quark mass which is
a function of $\lambda$.

Thus, we have derived in this section the bound state equations in the
weak-coupling scheme of the non-perturbative QCD for the light-light,
heavy-heavy and heavy-light mesons. By solving these equations
and comparing with experimental data, such as meson mass
spectroscopy, we can determine the $\lambda$ dependence of the
constituent quark masses, the effective coupling constant as
well as the wavefunction renormalization (anomalous dimensions)
of hadronic states. Then we are able to use the corresponding
wavefunctions to describe and predict various hadronic processes.
The low energy cutoffs, or more precisely, the low energy scale dependences
indeed reveal the inherent QCD dynamics of hadronic bound states.
For completeness, we should also derive the bound state equation for
glueball states ($gg$ bound states).  The glueball bound state
equation is not only the basis for the study of the currently
searching glueball states, but also allow us to determine
another very important quantity in the present weak-coupling
treatment of low energy QCD, i.e., the constituent gluon
mass and its scale dependence. But in this paper, we shall mainly
consider heavy hadron systems. As we have seen, in the heavy
quarkonium bound states, the constituent light quark and
gluon masses do not appear. We only need to determine the
effective coupling constant. Thus, the quarkonium states
are the simplest systems in present theory. In fact, the determination
of the scale dependence of the effective coupling constant $g_\lambda$
is the most important problem, from which we can test whether
a weak-coupling treatment of nonperturbative QCD can be
realized in this framework. In the following sections, we
shall study heavy quarkonia in details.  The extension to
heavy-light quark systems will be briefly discussed and the
more explicit study will be presented in the forthcoming
publication.

Before proceeding to solve the bound state equations derived
in this section, we shall demonstrate first from these bound state
equations how quark confinement is realized on the light-front.

\section{Quark confinement on the light-front}

In the conventional picture, QCD has a complex vacuum that
contains infinite quark pairs and gluons necessary
for confinement and  chiral symmetry breaking.
On the light-front, the longitudinal momentum of physical
particles is always positive, $p^+ = p^0 + p^3 \geq 0$.
As a result, only these constituents with zero longitudinal
momentum (called zero modes) can occupy the light-front vacuum.
The zero modes carry an extremely high light-front
energy which has been integrated out in the similarity
renormalization group scheme.  Consequently, some equivalent
effective interactions associated with the effect of the
nontrivial QCD vacuum are generated in the low energy
Hamiltonian $H_\lambda$.  Furthermore, the use of the constituent
picture in the weak-coupling scheme forbids possible occurrence of
any zero modes in $H_\lambda$ in the subsequent computations.
Therefore, light-front QCD vacuum remains trivial.
The nature of nontrivial QCD vacuum structure,
the confinement as well as the chiral symmetry breaking
must be made manifestly in $H_\lambda$ in terms of new
effective interactions. In fact, upon to the second
order calculation in the similarity renormalization scheme
(see sections II and III), the effective Hamiltonian
$H_\lambda$ already contains a confining interaction.
The interactions associated with the chiral symmetry breaking
may be manifested in the fourth order computation of $H_\lambda$,
as pointed out by Wilson \cite{Wilson94a}, but these
interactions are not important in the study of
 heavy hadrons.  Hence, next we shall
only discuss the quark confinement in terms of the
light-front bound state equations, from which a light-front
picture of confinement mechanism becomes transparent.

To be specific, we take the following criteria
as a definition of quark confinement: i) No color
non-singlet bound states exist in nature, only color singlet
states with finite masses can be produced and observed;
ii) There is a confining potential for quark interaction such that
quarks cannot be well-separated; iii) The conclusions of i--ii)
are only true for QCD but not for QED.  If conditions
i--iii) could be verified from the low energy effective QCD
Hamiltonian $H_{\lambda 0}$ and the corresponding bound state
equations, then quark confinement is realized on
the light-front. Here we shall take heavy quarkonia
as an explicit example. Some of the ideas for this confinement
picture have been discussed in \cite{Perry94,Wilson94a}.

In the present formulation of low-energy QCD,
non-existence of color non-singlet bound states is
essentially related to infrared divergences in the
effective Hamiltonian.
First of all, we shall show how only for physical states
the infrared divergence in the quark self-energy correction is
cancelled exactly by the same divergence from the uncancelled
instantaneous interaction in eq.(\ref{QQbse}). It is obvious
that when $y' \rightarrow y$, the uncancelled
instantaneous interaction leads to a severe infrared
divergence.  Assuming
that $\phi_{Q\overline{Q}}(y, \kappa_\bot)$ is a smooth
function with respect to $y$ and $\kappa_\bot$, and vanishes
when $y' \rightarrow \infty$. Then the dominant contribution
of the following integral is given by,
\begin{eqnarray}
	-4g_\lambda^2 && (T^a)(T^a) \int {dy' d^2\kappa'_{\bot}
		\over 2(2\pi)^3}{1 \over (y-y')^2} \theta( \lambda^2
		- A(y-y', \kappa_\bot - \kappa'_\bot, \overline{\Lambda}))
		\phi_{Q\overline{Q}}(y,\kappa'_{\bot}) \nonumber \\
	\sim && 4g_\lambda^2 (T^a)(T^a) \phi_{Q\overline{Q}}(y,
		\kappa_{\bot}) \int_{y -\epsilon}^{y+\epsilon}
		{dy' d^2\kappa'_{\bot}\over 2(2\pi)^3} {1 \over (y-y')^2}
		\theta( \lambda^2 - A(y-y', \kappa_\bot - \kappa'_\bot,
		\overline{\Lambda})) \nonumber \\
	&& = \Bigg({g_\lambda^2 \lambda^2 \over 2 \pi^2} (T^a)(T^a)
		\ln \epsilon \Bigg)~\phi_{Q\overline{Q}}(y,k_{\bot}) .
		\label{beid}
\end{eqnarray}
A more complete computation with an explicit light-front
wavefunction will be given in the next section.

Eq.(\ref{beid}) indicates that in bound state
equations, the uncancelled instantaneous interaction
contains a logarithmic infrared divergence.
Except for the color factor, this infrared divergence has the
same form as the divergence in the self-energy correlation.
{}From the bound state equation (\ref{QQbse}), we immediately
obtain the following conclusions.

	(a). For a single (constituent) quark state,
the bound state equation simply leads to
\begin{equation}
	\overline{\Lambda'}^2 = - {g_\lambda^2 \lambda^2 \over 4 \pi^2}
		C_f \ln \epsilon .
\end{equation}
This means that mass correction for single quark states
is infinite (infrared divergent) and cannot be renormalized
away in the spirit of gauge invariance.  Equivalently speaking,
single quark states carry an infinite mass and therefore
they cannot be produced.

	(b). For color non-singlet composite states, the color
factor $(T^a)_{\alpha \beta} (T^a)_{\delta \gamma}$ in the
$Q\overline{Q}$ interaction is different from the color factor
$\delta_{\alpha \beta} C_f=(T^a T^a)_{\alpha \beta}$.
Therefore, the infrared divergence in the self-energy correction
also cannot be cancelled by the corresponding divergence
from the uncancelled instantaneous interaction. As a result,
 color non-singlet composite states are infinitely heavy
so that they cannot be produced as well.

	(c). For color singlet $Q\overline{Q}$ states, the
color factor $(T^a)(T^a)$ becomes $(T^a T^a)= C_f$.
Thus, the infrared divergences in
eq.(\ref{QQbse}) are completely cancelled and the binding
energy of the corresponding bound states is guaranteed to
be finite.  In other words, only color singlet composite
particles are physically observable bound states, as a solution of
eq.(\ref{QQbse}).

The above conclusion is also true for heavy-light and
light-light hadronic states [see eqs.(\ref{lfbse}) and
(\ref{Qqbse})]. This provides the first condition for
quark confinement in light-front QCD.  Indeed, the physical
origin of the above result is very clear.  Light-front
infrared divergences are associated with violation of gauge
invariance. Only in gauge noninvariant sectors,
light-front infrared divergences may occur.  In gauge
invariant sectors, infrared divergences must be
automatically cancelled.   Therefore, the above conclusion
is a natural consequence of gauge invariance.

Physically, in order to be consistent with the above
conclusion, confinement must also imply the existence
of a confining potential so that quarks cannot be
well-separated to become asymptotically free states. Now we can
show that {\it the interactions in effective Hamiltonian (\ref{QQeh})
contains indeed both a confining and a Coulomb potentials.}

The Coulomb potential can be easily obtained by applying the
Fourier transformation to the second term in (\ref{QQvv}).
It is convenient to perform the calculation in the frame
$K_\bot =0$, in which
\begin{eqnarray}
	\overline{\Lambda}^2 && {1 \over (\kappa_{\bot} -
		\kappa_{\bot}')^2 + (y-y')^2 \overline{\Lambda}^2}
			\nonumber \\
	&& \sim  {1 \over 4\pi} \int dx^- d^2x_\bot e^{i(x^-q^+
		+ q_\bot \cdot x_\bot)}
		\Bigg({\overline{\Lambda} \over K^+} \Bigg) {1 \over
		\sqrt{x^2_\bot + \Big({\overline{\Lambda} \over K^+}
		\Big)^2 (x^-)^2}} ,	\label{Coulp}
\end{eqnarray}
where $q^+ = k_1^+-k_3^+=K^+(y-y'), q_\bot = k_{1\bot} - k_{3\bot} =
\kappa_\bot - \kappa'_\bot$ for $K_\bot =0$. Eq.(\ref{Coulp})
shows that the Coulomb potential on the light-front for quarkonium
states has the form
\begin{equation}
	V_{Coul.}(x^-, x_\bot) = - {g_\lambda^2 \over 4\pi} C_f
		{\overline{\Lambda} \over K^+} {1 \over \sqrt{x^2_\bot +
		\Big({\overline{\Lambda} \over K^+}\Big)^2 (x^-)^2}}
	 = - {g_\lambda^2 \over 4\pi} C_f {\overline{\Lambda} \over K^+}
		{1 \over r_l} ,
\end{equation}
where
\begin{equation}
	r_l = \sqrt{x^2_\bot + \Big({\overline{\Lambda}
		\over K^+} \Big)^2 (x^-)^2}
\end{equation}
is defined to be a ``radial'' variable in light-front
space\cite{Wilson94}.

The confining potential corresponds to the finite part of
the non-cancelled instantaneous interaction
in (\ref{QQvv}).  Its Fourier transformation is relatively
complicated.  The general expression is
\begin{eqnarray}
	 \int {dq^+ d^2q_\bot \over (2\pi)^2} && e^{i(q^+ x^-
		+ q_\bot \cdot x_\bot)} \Bigg\{-{4g_\lambda^2 C_f
		\over K^{+2}}~{1 \over (y-y')^2} \theta(\lambda^2
		- A(y-y', \kappa_\bot - \kappa'_\bot,
		\overline{\Lambda})) \Bigg\}  \nonumber \\
	&&= - {g_\lambda^2 \over 2\pi^2} C_f\int_0^{{\lambda^2 \over
		\overline{\Lambda}^2} K^+} dq^+ e^{iq^+ x^-}{q^2_{\bot m}
		\over q^{+2}}~{2J_1(|x_\bot|q_{\bot m}) \over
		|x_\bot| q_{\bot m}},		\label{conp}
\end{eqnarray}
where $q_{\bot m} = \sqrt{{\lambda^2 \over K^+} q^+ -
{\overline{\Lambda}^2 \over K^{+2}} q^{+2}}$, and $J_1 (x)$ is
a Bessel function.   An analytic solution to the
integral (\ref{conp}) may be difficult to carry out.
However, the nature of confining interactions is a large
distance QCD dynamics.  We may consider the integral
for large $x^-$ and $x_\bot$.  In this case, if $q^+ x^-$ and/or
$|x_\bot|q_{\bot m}$ are large, the integration vanishes,
yet $J_1(x) = {x\over 2} + {x^3\over 16} + \cdots$ for
small $x$.  The dominant contribution of the integral
(\ref{conp}) for large $x^-$ and $x_\bot$ comes from the small
$q^+$ such that $q^+ x^-$ and/or $|x_\bot|q_{\bot m}$ must
remain small and therefore
\begin{equation}
	 e^{iq^+ x^-}{2J_1(|x_\bot|q_{\bot m}) \over
		|x_\bot| q_{\bot m}} \sim 1 .
\end{equation}
This corresponds to
\begin{equation}
	q^+ < {1 \over x^-}~~~ {\rm and/or} ~~~~
		q^+ < {K^+ \over |x_\bot|^2 \lambda^2 } .
\end{equation}

If $q^+ < {1 \over x^-}< { K^+ \over |x_\bot|^2 \lambda^2 }$,
eq.(\ref{conp}) is reduced to
\begin{equation}
	-{g_\lambda^2 \over 2\pi^2} C_f \int_0^{1\over x^-} dq^+
		{1\over q^{+2}}\Bigg({\lambda^2 \over K^+} q^+ -
		{\overline{\Lambda}^2 \over K^{+2}} q^{+2} \Bigg)
	    = {g_\lambda^2\lambda^2\over 2\pi^2 K^+}
		C_f \Big(\ln |x^-| + \ln \epsilon \Big) , \label{conp1}
\end{equation}
where a term $\sim {1 \over x^-}$ is neglected since $x^-$ is large,
and $\epsilon$ is an infrared cutoff on $q^+$. The infrared
logarithmic divergence ($\sim \ln \epsilon$)
exactly cancels the divergence in the self-energy corrections
in $H_\lambda$, so that the remaining is a
logarithmic confining potential:
\begin{equation}
	V_{conf.}(x^-, x_\bot) =  {g_\lambda^2 \lambda^2\over
		2\pi^2 K^+} C_f \ln |x^-|.
\end{equation}
Similarly, when  $q^+ < { K^+ \over |x_\bot|^2 \lambda^2 }<
{1 \over x^-}$, we have
\begin{equation}
	-{g_\lambda^2 \over 2\pi^2} C_f \int_0^{K^+ \over
		|x_\bot|^2 \lambda^2} dq^+ {1\over q^{+2}}\Big({\lambda^2
		\over K^+} q^+ - {\overline{\Lambda}^2 \over
		K^{+2}} q^{+2} \Big) = {g_\lambda^2 \lambda^2\over
		2\pi^2 K^+} C_f \Big(\ln { \lambda^2 |x_\bot|^2 \over
		K^+} + \ln \epsilon \Big) ,	\label{conp2}
\end{equation}
where the term $\sim {1 \over x^2_\bot}$ has also been ignored
because of large $x^2_\bot$. Again, the infrared divergence
($\sim \ln \epsilon$) is cancelled in $H_\lambda$, and we obtain the
following confining potential:
\begin{equation}
	V_{conf.}(x^-, x_\bot) =  {g_\lambda^2 \lambda^2\over 2\pi^2 K^+}
		C_f \ln {\lambda^2 |x_\bot|^2 \over K^+}.
\end{equation}
Thus, the effective Hamiltonian $H_{\lambda 0}$ contains a logarithmic
confining potential in all the directions of $x^-$ and $x_\bot$
coordinates. Note that for heavy quarkonia, a logarithmic confining
potential provides indeed a good description to the spectroscopy
and leptonic decays \cite{Gigg}. More details of computation
will be given in the next section. Nevertheless, we have explicitly
shown here that $H_{\lambda 0}$ exhibits a Coulomb potential at
short distances and a confining potential at long distances.
The second condition for quark confinement
is verified on the light-front.

Finally, we shall argue that the above mechanism of quark
confinement is indeed only true for QCD.  As we have
seen the light-front confinement potential is just an
effect of the non-cancellation between instantaneous
interaction and one transverse gluon interaction.  Such
a non-cancellation arises from
the requirement in the similarity renormalization group
scheme that the transverse gluon energy cannot be below
a certain value (about a few MeV). This requirement is
naturally satisfied if the gluon mass is nonzero in low
energy scale.  Unlike the constituent quark mass which we know
is an effect of the spontaneous chiral symmetry breaking,
the origin of constituent massive gluons is not very clear at
present. The assumption of massive gluons here may also violate
gauge invariance but it is not unnatural.  In fact,
if gluon were massless like photons, the hadronic spectra
would be continuous rather than discrete, as Wilson
pointed out recently\cite{Wilson94a}. A typical evidence of
gluons being massive in the low energy domain is the possible
existence of glueball states which is still a very
active topic in current experimental searches \cite{Glueball}.
The massive gluon must be originated from the nonlinear interactions
in non-abelian gauge theory. Therefore, the non-cancellation
of the instantaneous interaction in the low energy domain
is indeed a consequence of the existence of the constituent
massive gluons due to the non-abelian gauge interactions.  This
is independent of any particular renormalization scheme.
The use of the low energy cutoff $\lambda$ just gives us a simple
realization of this confining picture that the massive gluon exchange
energy cannot run down to the zero value in nonperturbative QCD.
In the case of lacking the mechanism of how the massive gluons
are generated, the determination of the gluon mass
lies on the solution of bound state equations.

Based on the above discussion, it is now easy to
find that the confinement mechanism described in this
section is not valid in QED.  First of all, the
infrared divergence in the self energy is also a result of
the noncancellation between the instantaneous interaction
self-energy diagram and the one-loop self-energy
corrections [see eq.(\ref{cqse})].  In QED, since the photon
mass is always zero, the photon energy in the one-loop
self-energy correction covers the entire range from zero to
infinity.  Thus, in QED, we can always choose the low energy
cutoff $\lambda$ being zero. (We shall further explain
in Section VII that this is indeed the only choice
for applying similarity renormalization group approach
to QED. Otherwise the resulting effective QED theory is
inconsistent with the perturbative QED theory.) Then a
direct calculation for (\ref{cqse}) with $\lambda=0$
shows that the infrared
divergences do not occur in the electron self-energy
correction.  As a result, the renormalized single electron mass
is finite, in contrast to the divergent mass of single quark
states.  For the same reason, with $\lambda=0$,
the instantaneous interaction in
the effective QED Hamiltonian is also exactly cancelled
by the same interaction from one transverse photon exchange
so that only one photon exchange Coulomb interaction remains.
Therefore, using similarity renormalization group approach to
QED, we obtain a conventional effective QED Hamiltonian which
only contains the Coulomb interaction.  Such an effective
Hamiltonian is the basis in the study of positronium
bound states.  More discussion will be given in Section VII.

Now we shall study how a weak-coupling treatment scheme
works in solving hadronic bound state problem in the present
low energy QCD formulation.

\section{QCD description of quarkonia on the light-front}

A numerical computation to the heavy hadron bound state equations,
eqs.(\ref{QQbse}) and (\ref{Qqbse}), is actually not too difficult.
However, to have a deeper insight about the internal structure of
light-front bound states and to determine the scale dependence
of the effective coupling constant in $H_\lambda$, it is better to have
an analytic analysis.  In this paper, the light-front wavefunction
ansatz will be used to solve the bound state equations for heavy
quarkonia, from which some general properties of the low energy
scaling in the similarity renormalization group can be extracted.
It also provides a direct test whether the weak-coupling
treatment of nonperturbative QCD can be realized.

\subsection{A general analysis of light-front wavefunctions}

For heavy quark systems, the wavefunctions considered in
the previous section are defined in heavy mass limit.
Most of the $1/m_Q$ corrections can be handled in the standard
perturbation theory in the present framework, except
for the kinetic energy for quarkonia.  The heavy
hadronic wavefunctions in the heavy mass limit
can be tremendously simplified.

First of all, the heavy quark kinematics have already added
some constraints on the general form of the light-front
wavefunction $\phi (x, \kappa_\bot)$.  The kinetic energy part (the
left hand side of these bound state equations in section IV)
shows that when we introduce the residual longitudinal
momentum fraction $y$ for heavy quarks, the longitudinal
momentum fraction dependence in $\phi$ is quite different
for the heavy-heavy, heavy-light and light-light mesons.

For the light-light mesons, such as pions, rhos, kaons etc.,
the wavefunction $\phi_{q\overline{q}}
(x,\kappa_\bot)$ must vanish at the endpoint $x=0$ or $1$.
This can be seen from the kinetic energy
contribution in the bound state equation [see eq.(\ref{lfbse})],
\begin{equation}	\label{kc1}
	M^2 - M_0^2 = M^2 - {\kappa_\bot^2 + m_1^2 \over x}
		- {\kappa_\bot^2 + m_2^2 \over 1-x} .
\end{equation}
To ensure that the bound state equation is well defined
in the entire range of momentum space, $|\phi_{q\overline{q}}
(x,\kappa_\bot)|^2$ must fall down to zero in the
longitudinal direction not slower than $1/x$
and $1/(1-x)$ when $x \rightarrow 0$ and $1$, respectively.
In other words, at least $\phi_{q\overline{q}}(x,\kappa_\bot)
 \sim \sqrt{x(1-x)}$ .

For heavy-light quark mesons, namely the $B$ and $D$ mesons,
the wavefunction $\phi_{Q\overline{q}}(y,\kappa_\bot)$ is
required to vanish at $y=0$, where $y$ is the residual
longitudinal momentum fraction carried by the light quark.
This is because the kinetic energy in
eq.(\ref{Qqbse}) contains a singularity at $y=0$,
\begin{equation}
	M^2 - M_0^2  \longrightarrow \overline{\Lambda}^2
		- \overline{M}_0^2 = \overline{\Lambda}^2
		+ (1-y) \overline{\Lambda}^2 - {\kappa_\bot^2
		+ m_1^2 \over y}.	\label{kc2}
\end{equation}
On the other hand, since  $0 \leq y \leq \infty$,
$\phi_{Q\overline{q}}(y,\kappa_\bot)$ should also vanish
when $y \rightarrow \infty$.  Hence, a possible simple form is
$\phi_{Q\overline{q}}(y,\kappa_\bot) \sim \sqrt{y}e^{-\alpha y}$
or $  \sqrt{y} e^{-\alpha y^2}$.  The $y$ dependence
in $\phi_{Q\overline{q}}(y,\kappa_\bot)$ is obviously different
from the $x$ dependence in $\phi_{q\overline{q}}(x,\kappa_\bot)$.

For heavy quarkonia, the kinetic energy in the corresponding
bound state equation (\ref{QQbse}) is:
\begin{equation}	\label{kc3}
	M^2 - M_0^2  \longrightarrow \overline{\Lambda}^2
		- \overline{M}_0^2 = 2\overline{\Lambda}^2
		-{\overline{\Lambda}\over m_Q}\Big[2\kappa_\bot^2
		+ \overline{\Lambda}^2(2y^2 -2y +1)\Big].
\end{equation}
Since $-\infty < y < \infty$, the normalization forces
$\phi_{Q\overline{Q}}(y,\kappa_\bot)$ to vanish as
$y \rightarrow \pm \infty$. Therefore a possible form is
$\phi_{Q\overline{Q}}(y,\kappa_\bot) \sim e^{-\alpha y^2}$.
Obviously, in heavy quark mass limit, the $y$ dependence
in $\phi_{Q\overline{Q}}(y,\kappa_\bot)$ is very
different from the above two cases.

On the other hand, the transverse momentum dependence in
these light-front wavefunctions should be more or less
similar.  They all vanish at $\kappa_\bot
\rightarrow \pm \infty$. A simple form of the $\kappa_\bot$
dependence for these wavefunctions is a Gaussian function:
$e^{- \kappa_\bot^2 / 2\omega^2}$.

The above analysis of light-front wavefunctions is only
based on the kinetic energy properties of the constituents.
Currently, many investigations on the hadronic
structures use phenomenological light-front wavefunctions.
One of such phenomenological wavefunctions that has been
widely used in the study of heavy hadron structure is the so-called
BSW wavefunction, introduced by Bauer et al.\cite{BSW},
\begin{equation}	\label{bsw}
	\phi_{BSW}(x,\kappa_\bot) = {\cal N} \sqrt{x(1-x)}
		~\exp\left(-{\kappa^2_\bot\over2\omega^2}\right)
		~\exp\left[-{M_H^2\over2\omega^2}(x-x_0)^2\right],
\end{equation}
where ${\cal N}$ is a normalization constant, $\omega$
a parameter of order $\Lambda_{QCD}$, $x_0=({1\over2}
-{m_1^2-m_2^2\over2M_H^2})$, and $M_H$, $m_1$, and $m_2$ are
the hadron, quark, and antiquark masses respectively. In the
phenomenological description, the parameters $\omega$, and
$m_i~(i=1,2)$ in (\ref{bsw}) are fitted from data. Here
we are of course interested in the dynamical determination
of these parameters.

As we have pointed out in passing, for heavy quark systems, the
$1/m_Q$ corrections can be well treated perturbatively
in our framework (except for the kinetic energy of
quarkonia).  Here we are only interested in the solution of
the wavefunctions in the heavy mass limit,
where eq.(\ref{bsw}) can be further simplified.

Explicitly, for heavy-light quark systems, such as the $B$ and $D$
mesons, one can easily find that in the heavy mass limit,
\begin{equation}
	m_1 = m_Q \sim M_H, ~~m_q << m_Q ~~~ {\rm so~that}~~~ x_0 =0 .
\end{equation}
Meanwhile, from eq.(\ref{Qqlm}), we also have
\begin{equation}
	M_H x = \overline{\Lambda} y.
\end{equation}
Furthermore, the factor
$\sqrt{x(1-x)}$ can be rewritten by $\sqrt{y}$ in according to
the discussion on eq.(\ref{kc2}). Thus, the BSW wavefunction
is reduced to
\begin{equation}	\label{Qqtwf}
	\phi_{Q\overline{q}}(y,\kappa_\bot) = {\cal N} \sqrt{y}
		~\exp\left(-{\kappa_\bot^2\over2\omega^2}\right)
	~\exp\left(-{\overline{\Lambda}^2\over2\omega^2}y^2\right).
\end{equation}
This agrees with our qualitative analysis given
before. Using such a wavefunction we
have already computed the universal Isgur-Wise function in
$B \rightarrow D, D^*$ decays \cite{Cheung95},
\begin{equation}
	\xi( v \cdot v') = {1 \over v \cdot v'},   \label{iwf}
\end{equation}
and from which we obtained the slope of $\xi(v \cdot v')$
at the zero-recoil point, $\rho^2 = - \xi'(1) = 1$,
in excellent agreement with the recent CLCO result \cite{clco}
of $\rho^2 = 1.01 \pm 0.15 \pm 0.09$. This result is
independent of the value of $\omega$, and therefore is
independent of further dynamics of QCD involved in
the corresponding bound state equation. The simple
form (\ref{Qqtwf}) is just a consequence
of heavy quark symmetry in the our low energy theory.
We may argue that $\rho^2=1$ could be a universal
identity.

For heavy quarkonia, such as the $b\overline{b}$ and $c\overline{c}$
states, $m_1 = m_2 = m_Q$ which leads to $x_0=1/2$ in eq.(\ref{bsw}).
Also, the longitudinal momentum fraction in (\ref{bsw}) is defined
by $x=p_1^+/P^+$, its relation to the residual longitudinal
momentum fraction is given by
\begin{equation}
	M_H (x- {1\over 2}) = \overline{\Lambda} y .	\label{xyr}
\end{equation}
In addition, the factor
$\sqrt{x(1-x)}$ must be totally dropped as we have seen from
the discussion on eq.(\ref{kc3}). Therefore the BSW wavefunction
for quarkonia is reduced to
\begin{equation}	\label{QQtwf1}
	\phi_{Q\overline{Q}}(y,\kappa_\bot) = {\cal N} ~\exp \left(
		-{\kappa^2_\bot\over2\omega^2}\right) ~\exp\left(
		-{\overline{\Lambda}^2\over2\omega^2}y^2\right),
\end{equation}
which is the exact form as we expected from the qualitative
analysis. Here we have not taken the limit of $m_Q
\rightarrow \infty$ for heavy quarkonia. Thus a possible $m_Q$
dependence in wavefunction may be hidden in the parameter $\omega$.

Another phenomenological light-front wavefunction
which has been widely used for both light and heavy
mesons has the form \cite{Chung}
\begin{equation}
	\psi_{q\overline{q}}(x, \kappa_\bot) ={\cal N}
		\sqrt{{d\kappa_z\over dx}} ~\exp\left(-{\kappa_\bot^2
		\over 2\omega^2}\right)~\exp\left(-{\kappa_z^2
		\over 2\omega^2}\right),  \label{glf}
\end{equation}
where $\kappa_z$ is defined by
\begin{equation}
	x = {e_1 + \kappa_z \over e_1 + e_2}, ~~~ 1-x = {e_2 -\kappa_z
		\over e_1 + e_2}; ~~~ e_i = \sqrt{\kappa_\bot^2
		+ \kappa_z^2 + m_i^2}~~(i=1,2)
\end{equation}
as a pretended $z$-component of relative momentum while
 $\sqrt{d\kappa_z/dz}$ is the Jacobian of transformation from
$(x, \kappa_\bot)$ to $\vec \kappa =(\kappa_\bot, \kappa_z)$.
This wavefunction has been used frequently in various
studies of hadronic transitions. In particular, it has been
shown that this wavefunction describes satisfactorily the pion
elastic form factor up to $Q^2\sim 10 ~{\rm GeV}^2$ \cite{Chung}.

For heavy quarkonia with $m_1 = m_2 = m_Q$, we may have
\begin{equation}
	\kappa_z = M_0 (x - {1 \over 2}) \longrightarrow
		\overline{\Lambda}y .
\end{equation}
Thus, $\sqrt{d\kappa_z /dx} =$ constant, and eq.(\ref{glf})
is reduced to the same form obtained from the BSW wavefunction,
\begin{equation}
	\phi_{q\overline{q}}(x,\kappa_\bot) \longrightarrow
		\phi_{Q\overline{Q}}(y,\kappa_\bot) = {\cal N} ~\exp
		\left(-{\kappa^2_\bot\over2\omega^2}\right) ~\exp
		\left(-{\overline{\Lambda}^2\over2\omega^2}y^2\right).
\end{equation}
Therefore, the above Gaussian-type ansatz should be a very
good candidate for the low-lying quarkonium states.
In the following, we start with this wavefunction ansatz
to solve the light-front quarkonium bound state equation,
and from which to determine the low energy scaling dynamics
and develop the weak-coupling treatment of the heavy hadron
bound states.

\subsection{A weak-coupling realization of
the nonperturbative QCD description for heavy quarkonia}

Based on the analysis in the last section, we take the normalized
wavefunction ansatz of (\ref{QQtwf1}),
\begin{equation}
	\phi_{Q\overline{Q}}(y,\kappa_\bot) = 4\sqrt{\overline{
		\Lambda}} \Bigg({\pi \over \omega_\lambda^2}\Bigg)^{3/4}
		\exp \Bigg(-{\kappa_\bot^2 \over 2 \omega_\lambda^2} \Bigg)
		\exp \Bigg( - { \overline{\Lambda}^2 \over 2
		\omega_\lambda^2} y^2 \Bigg) ,	\label{QQtwf2}
\end{equation}
as a solution (a trial wavefunction) of the heavy quarkonium
bound state equation (\ref{QQbse}).
Note that here we have also specified the scale dependence of the
wavefunction through the scale dependence of the parameter
$\omega_\lambda$. Substituting the above wavefunction into
the quarkonium bound state
equation (\ref{QQbse}) and introducing the new variables
\begin{eqnarray}
	&& Z = {1\over 2}(y + y'),~~~ z = y - y' , \nonumber \\
	&& Q_\bot = {1\over 2}(\kappa_\bot + \kappa'_\bot), ~~~
		q_\bot = \kappa_\bot - \kappa'_\bot,
\end{eqnarray}
we have
\begin{eqnarray}
	2\overline{\Lambda}^2 &=& {\overline{\Lambda}\over m_Q}
		\Big(3\omega_\lambda^2 + \overline{\Lambda}^2\Big)
		- {g_\lambda^2\over 2 \pi^2}\lambda^2 C_f
		\ln\epsilon \nonumber \\
	&& ~~~~~  -4g_\lambda^2 C_f \int {dz d^2q_{\bot} \over 2(2\pi)^3}
		\exp \Bigg\{-{1 \over 4 \omega_\lambda^2}( q_\bot^2 + z^2
		\overline{\Lambda}^2) \Bigg\} \nonumber \\
	&& ~~~~~~~~~~~~~~~~~ \times
		\Bigg\{ {1 \over z^2} \theta(\lambda^2|z| -  q_\bot^2 - z^2
		\overline{\Lambda}^2) + {\overline{\Lambda}^2
		\over q_\bot^2 + z^2 \overline{\Lambda}^2 } \theta(
		q_\bot^2 + z^2 \overline{\Lambda}^2 - \lambda^2|z|) \Bigg\}
		  	\nonumber \\
	&=& {\cal E}_{kin} - {g_\lambda^2\over 2 \pi^2}
		\lambda^2 C_f \ln\epsilon + {\cal E}_{nonc} +
		{\cal E}_{Coul} , \label{QQev}
\end{eqnarray}
where ${\cal E}_{kin}$ represents the kinetic energy,
${\cal E}_{nonc}$ is the contribution of the noncancellation
of the instantaneous interaction, and ${\cal E}_{Coul}$
from the Coulomb interaction. Furthermore, it is not very
difficult to compute that
\begin{eqnarray}
{\cal E}_{nonc} &=& -4g_\lambda^2 C_f \int {dz d^2q_{\bot} \over
		2(2\pi)^3}{1 \over z^2} \theta(\lambda^2|z| -
		q_\bot^2 - z^2 \overline{\Lambda}^2) \exp
		\Bigg\{-{1 \over 4 \omega_\lambda^2}
		(q_\bot^2 + z^2 \overline{\Lambda}^2) \Bigg\} \nonumber \\
	&=& {g_\lambda^2\over 2\pi^2} C_f \lambda^2 \Bigg\{ \gamma +
		\ln {\lambda^2 \epsilon \over 4\omega^2_\lambda}
		+  {\rm E}_1(\varpi^2) + {\sqrt{\pi}\over \varpi}
		{\rm Erf}(\varpi) \Bigg\},  	\label{confe} \\
 {\cal E}_{Coul} &=& -4g_\lambda^2 C_f \int {dz d^2q_{\bot}
		\over 2(2\pi)^3} {\overline{\Lambda}^2 \over q_\bot^2
		+z^2\overline{\Lambda}^2}
		 \theta( q_\bot^2 +z^2\overline{\Lambda}^2 - \lambda^2|z|)
		\exp \Bigg\{-{1 \over 4 \omega_\lambda^2}( q_\bot^2+z^2
		\overline{\Lambda}^2) \Bigg\} \nonumber \\
	&=& - {g_\lambda^2\over 2\pi^2} C_f
		\lambda^2\Bigg\{ {\sqrt{\pi}\over \varpi}
		\Big[1-~{\rm Erf}(\varpi)\Big] + {1 \over \varpi^2}
		\Big[1 - e^{-\varpi^2} \Big] \Bigg\} ,
\end{eqnarray}
where  $\gamma=0.57721566...$ is the Euler constant, $\epsilon$
is the small longitudinal momentum cutoff, the dimensionless
$\varpi$ is defined by
\begin{equation}
	\varpi = {\lambda^2 \over 2 \omega_\lambda \overline{\Lambda}},
\end{equation}
and E$_1$ and Erf are the exponential integral function and the
error function, respectively,
\begin{equation}
	{\rm E}_1 (x) = \int_x^\infty { e^{-t} \over t}  dt~~, ~~~~
		{\rm Erf}(x) = {2 \over \sqrt{\pi}} \int_0^x e^{-t^2} dt .
\end{equation}
We may rewrite the term $\ln {\lambda^2 \epsilon \over 4
\omega^2_\lambda}$ in ${\cal E}_{nonc}$ as
\begin{equation}
	\ln {\lambda^2 \epsilon \over 4\omega^2_\lambda} = \ln
		\epsilon + \ln \varpi^2 + \ln {\overline{\Lambda}^2
		\over \lambda^2} .
\end{equation}
It shows that ${\cal E}_{nonc}$ contains a logarithmic
divergence $\ln \epsilon$ which exactly cancels the same divergence
from the self-energy correction, as expected, and the term
$\ln \varpi^2$ is the logarithmic confining energy.

After the cancellation of the infrared $\ln \epsilon$ divergences
in eq.(\ref{QQev}), the binding energy for heavy quarkonia is
given by the kinetic energy plus the potential energy:
\begin{eqnarray}
	2\overline{\Lambda}^2 &=& {\cal E}_{kin} + {\cal E}_{conf}
		+ {\cal E}_{Coul} \nonumber \\
	 &=& {\overline{\Lambda} \over m_Q} \Bigg\{3\omega_\lambda^2
		+ \overline{\Lambda}^2 \Bigg\} +
		{g_\lambda^2\over 2\pi^2}C_f \lambda^2 \Bigg\{
		\gamma + \ln \varpi^2 + \ln {\overline{\Lambda}^2
		\over \lambda^2} + E_1(\varpi^2) + {\sqrt{\pi}
		\over \varpi}{\rm Erf}(\varpi) \Bigg\}  \nonumber \\
	& & ~~~~~~~~~~~~~~~~
		- {g_\lambda^2\over 2\pi^2}C_f \lambda^2
		\Bigg\{ {\sqrt{\pi} \over \varpi}
		\Big[1- {\rm Erf}(\varpi)\Big]+{1 \over \varpi^2}
		\Big[1-e^{-\varpi^2}\Big] \Bigg\} \nonumber \\
	&=& {\overline{\Lambda} \over m_Q} \Bigg\{3\omega_\lambda^2
		+ \overline{\Lambda}^2 \Bigg\} + {g_\lambda^2\over
		2\pi^2}C_f \lambda^2 \Bigg\{F(\varpi)
	    +\ln {\overline{\Lambda}^2 \over \lambda^2} \Bigg\},
		\label{sQQe}
\end{eqnarray}
where
\begin{equation}	\label{dlsf}
	F(\varpi) = \gamma + \ln \varpi^2 + E_1(\varpi^2)
		- {\sqrt{\pi} \over \varpi}\Big[1 -
		2{\rm Erf}(\varpi)\Big]-{1 \over \varpi^2}
		\Big[1-e^{-\varpi^2}\Big] ,
\end{equation}
which is a dimensionless function. In Fig.1, we plot the confining
potential energy, the Coulomb potential energy and the totally
potential energy as functions of $\varpi$ which is proportional
to the radial variable in light-front space,
\begin{equation}
	\varpi \sim {1 \over \omega_\lambda} \sim r_l .
\end{equation}
Fig.1 shows that {\it the total potential energy is
a typical combination of the Coulomb potential in short
distance and a confining potential in long distance
that has been widely used in previous phenomenological
describing hadronic states, but it is now explicitly derived
from QCD.}  Furthermore, eq.(\ref{sQQe}) also indicates
that without considering the kinetic energy, we cannot
find stable quarkonium bound states.  The kinetic energy
balances the potential energy and ensures the
existence of a stable solution for (\ref{sQQe}).  Therefore,
the first order kinetic energy in HQET is an important
nonperturbative effect in binding two heavy quarks, as
noticed first by Mannel et al.\cite{Mannel95} in their
attempt of applying HQET to heavy quarkonium system.

If we know the experimental value of the quarkonium
binding energy $\overline{\Lambda}$, minimizing eq.(\ref{sQQe})
can completely determine the parameter $\omega_\lambda$
and the coupling constant $g_\lambda$.  The precise
value of quarkonium binding energy that can be compared with
the data in Particle Data Group \cite{PDB} must include
the spin-splitting energy ($1/m_Q$ corrections) which we
will present in the forthcoming paper \cite{Zhang96}.
Here, to justify whether a weak-coupling treatment
of the nonperturbative QCD can become possible
in the present formulation,
we will give a schematic calculation. It is known that
$\overline{\Lambda}$ is of the same order as  $\Lambda_{QCD}$
which is about $100 \sim 400$ MeV.  To solve (\ref{sQQe}) we
shall take several values of $\overline{\Lambda}$ within the
above range.  We choose the low energy cutoff about a typical
hadronic energy, $\lambda =1$ GeV. The charmed and bottom quark masses
used here are $m_c=1.4$ GeV and $m_b=4.8$ GeV. The results are
listed in Tables I and II for charmonium and bottomonium,
respectively, where $\omega_{\lambda 0}$ denotes the
minimum point of the binding energy (\ref{sQQe}).

We see from the Tables I--II that the coupling
constant $\alpha_\lambda = g^2_\lambda/4\pi$ is very small.
For instance, with $\overline{\Lambda} = 200$ MeV,
we obtain
\begin{equation}
	\alpha_\lambda = \left\{ \begin{array}{ll} 0.02665~~~~~~~
		& {\rm charmonium,} \\
		0.06795~~~~ & {\rm bottomonium,} \end{array} \right.
\end{equation}
which is much smaller than that extrapolated from the
running coupling constant in the naive perturbative
QCD calculation. The parameter $\omega_{\lambda 0}$
in the quarkonium wavefunction is the mean value
of the (transverse) momentum square of heavy quark
inside the heavy quarkonia:
\begin{equation}
	\langle k_\bot^2 \rangle = \omega_\lambda^2 .
\end{equation}
For charmonium, we can see that the resulting $\omega_\lambda$
are  typical values of $\Lambda_{QCD} \sim
\overline{\Lambda}$. The kinetic energy is about a half of the
potential energy.  For bottomonium, we find that the binding
energy $\overline{\Lambda}$ cannot be too large.  In fact, when
$\overline{\Lambda}$ is over about 260 MeV, eq.(\ref{sQQe})
has no solution.  Meanwhile, compared to charmonium, the
effective coupling constant is relatively large
(in contrast to the perturbative running coupling constant
which is smaller with increasing $m_Q$
if it is taken as the mass scale).  Also the values of
$\omega_\lambda$ in bottomonium wavefunctions are larger
than that in charmonium.  The difference between
charmonium and bottomonium in the nonperturbative
calculation may be understood as follows.  As we know,
in the nonrelativistic quark model, the quark momentum
in quarkonia is proportional to the quark mass,
$\omega_\lambda \sim m_Q$ \cite{Gigg}.  Our relativistic QCD
bound state solution exhibits such a property. This is why the
values of $\omega_\lambda$ for bottomonium are much
larger than that for charmonium. As a result, the bottomonium
kinetic energy ($\sim \omega_\lambda^2$) becomes large as well.
To have a nonperturbative balance between the kinetic energy and
the potential energy in the bound states, the coupling constant
in bottomonium must be larger than that in charmonium. All these
properties now have been manifested in the solution of eq.(\ref{sQQe}).
A more precise determination of $\alpha_\lambda$ (i.e.,
$g_\lambda$) requires an accurate computation of the low-lying
quarkonium spectroscopy with the $1/m_Q$ corrections
included\cite{Zhang96}.  Nevertheless, it has been shown that
the effective coupling constant in the low energy
Hamiltonian $H_\lambda$ is very small at the hadronic
mass scale.

In order to see how this weak coupling constant varies with the
cutoff $\lambda$, we take $\overline{\Lambda} =200$ MeV and
vary the value of $\lambda$ around 1 GeV. The result is listed
in Table III. We find that the coupling constant is decreased
very faster with increasing $\lambda$.  In other
words, {\it with a suitable choice of the low energy cutoff $\lambda$
in the similarity renormalization group scheme, we can make the
effective coupling constant $\alpha_\lambda$  in $H_\lambda$
arbitrarily small, and therefore the weak-coupling treatment of
the non-perturbative QCD can be achieved in terms of $H_\lambda$
such that the corrections from $H_{\lambda I}$ can be computed
perturbatively}.  Thus, we have provided the first
explicit realization of recently proposed
the weak-coupling treatment of nonperturbative QCD on the
light-front \cite{Wilson94}.

We must emphasize here that
$\alpha_\lambda$ is not the physical coupling constant
$\alpha_s = g_s^2/4\pi$.  The later is of order unity
at the hadronic mass scale. A detailed analysis of the
$\lambda$-dependence and the relation between $\alpha_\lambda$
and $\alpha_s$ will be discussed in the next.

\section{Similarity renormalization group equation
		and low energy running coupling constant}

In this section, we shall discuss the scale dependence
of the coupling constant, the constituent quark and gluon masses
as well as the wavefunctions.  For heavy quarkonia, the bound
state equation does not include the constituent gluon mass.
The heavy quark mass is larger than the usual hadronic mass
scale.  Its $\lambda$-dependence should be very weak that
can be neglected in the present discussion. Thus the remainings
are the $\lambda$-dependence of the coupling constant $g_\lambda$
and the wavefunctions, the later is described through the
$\lambda$-dependence of the parameter $\omega_\lambda$.

{}From these solutions in Tables I to III, we find that the values
of dimensionless parameter $\varpi={\lambda^2\over 2\overline{\Lambda}
\omega_{\lambda 0}}$ are greater than 2.5. When
$x > 2.5$, the exponential integral function and the error function
are simply reduced to E$_1(x)=0$ and Erf$(x)=1$. Thus,
the dimensionless function $F(\varpi)$ can be expressed
approximately by
\begin{equation}
	F(\varpi) = \gamma +  \ln \varpi^2 + {\sqrt{\pi} \over
		\varpi} -{1 \over \varpi^2}~~ , ~~~~ \varpi \geq 2.5,
\end{equation}
with an error less than $10^{-5}$. Hence we can simply rewrite
 eq.(\ref{sQQe}) as
\begin{equation}
	2\overline{\Lambda}^2 = {3\overline{\Lambda} \over m_Q}
		\omega_\lambda^2 - {g_\lambda^2\over 2\pi^2}C_f
		\lambda^2 \Bigg\{{4\overline{\Lambda}^2 \omega_\lambda^2
		\over \lambda^4} - \sqrt{\pi}~{2\overline{\Lambda}
		\omega_\lambda \over \lambda^2} - \Big(\gamma + \ln
		{\lambda^2 \over 4} -\ln \omega_\lambda^2\Big)~\Bigg\}
		+ {\overline{\Lambda}^3 \over m_Q} . \label{sQQe1}
\end{equation}
Minimizing $\overline{\Lambda}$ with respect to $\omega_\lambda$,
we obtain
\begin{equation}	\label{minimizing}
	 \Bigg\{ {3\overline{\Lambda} \over m_Q} - {g_\lambda^2 \over
		2\pi^2}C_f{4\overline{\Lambda}^2 \over \lambda^2}
		\Bigg\} \omega_{\lambda}^2 = {g_\lambda^2 \over
		2\pi^2}C_f \lambda^2 \Bigg\{ 1 - \sqrt{\pi}~
		{\overline{\Lambda} \omega_\lambda \over
		\lambda^2} \Bigg\}.
\end{equation}
Therefore, eq.(\ref{sQQe1}) becomes
\begin{equation}
	2\overline{\Lambda}^2 = {g_\lambda^2\over 2\pi^2}C_f \lambda^2
		\Bigg\{ 1+ \gamma + \ln {\lambda^2 \over 4} - \ln
		\omega_{\lambda 0}^2 + \sqrt{\pi}~{\overline{\Lambda}
		\omega_{\lambda 0}\over\lambda^2}~\Bigg\} +
		{\overline{\Lambda}^3 \over m_Q},  \label{sQQe2}
\end{equation}
where $\omega_{\lambda 0}$ is a solution of (\ref{minimizing}).
Eqs.(\ref{minimizing}) and (\ref{sQQe2}) determine the
$\lambda$-dependences of the coupling constant $g_\lambda$
and the wavefunction parameter $\omega_{\lambda 0}$.

Directly and analytically solving eqs.(\ref{minimizing}) and
(\ref{sQQe2}) is not obviously possible. The nonperturbative
balance between the kinetic energy and the potential energy
implies that $\omega_\lambda \sim \sqrt{m_Q \overline{\Lambda}}$.
Meanwhile, since it is the binding energy of heavy quarkonia,
$\overline{\Lambda}$ should be $\lambda$-independent.
We then obtain
\begin{eqnarray}
	\alpha_\lambda &=& {g^2_\lambda \over 4\pi} = {\pi \over
		C_f}~ \Bigg({\overline{\Lambda}^2 \over \lambda^2}
		\Bigg)\Bigg(1 - {\overline{\Lambda} \over 2m_Q}\Bigg)
		{1\over  1+ \gamma + \ln {\lambda^2 \over 4} - \ln
		\omega_{\lambda 0}^2 + \sqrt{\pi}~{\overline{\Lambda}
		\omega_{\lambda 0}\over\lambda^2 }} \nonumber \\
		&=& {\pi \over C_f}~ \Bigg( {\overline{\Lambda}^2
		\over \lambda^2}\Bigg)~{1\over a + b \ln {\lambda^2
		\over \overline{\Lambda}^2}}, \label{running}
\end{eqnarray}
where the coefficients $a$ and $b$ can be obtained by numerically
solving eqs.(\ref{minimizing}) and (\ref{sQQe2}). The coefficient
$b$ is almost a constant (with a slight dependence on $m_Q$
but independence on $\overline{\Lambda}$ and $\lambda$), while $a$
depends on $\overline{\Lambda}, \lambda$ and also $m_Q$.  For
$\lambda \geq 0.6$ GeV, the $\lambda$-dependence of the
parameter $a$ is negligible. In Fig.2, we plot the
$\lambda$-dependence of the effective coupling constant
$\alpha_\lambda$ for charmonium.  The dots are the numerical solutions
of (\ref{sQQe}) and the solid line is given by the analytical form
(\ref{running}) with $b=1.15$, and  $a=-0.25$ for
$\overline{\Lambda}=0.2$ GeV and $a=1.1$ for $\overline{\Lambda}
=0.4$ GeV. We can see that (\ref{running}) is a very good
analytical solution of the eqs.(\ref{minimizing}) and
(\ref{sQQe2}) [or of the minimizing eq.(\ref{sQQe})].

The above solution shows that with increasing $\lambda$,
$\alpha_\lambda$ becomes weaker and weaker. Meanwhile,
we also find that the confining energy becomes more
and more dominant in the binding energy (See Table IV).
Fig.3 is a plot of the parameter $\omega_{\lambda 0}$
as a function of $\lambda$, from which we also see that with
increasing $\lambda$, $\omega_{\lambda 0}$
is decreased. Correspondingly, the distance between
two quarks inside the quarkonia, $r_l \sim {1\over
\omega_{\lambda 0}}$,
is increased.  This is why the confining interaction
becomes more and more important.  On the other hand,
the confining interaction comes from the noncancellation
of instantaneous gluon exchange with energy below the
scale $\lambda$.  With the larger $\lambda$, the more
the instantaneous interaction contributes to $H_\lambda$.
Thus, the above conclusion can also be directly understood
from the low energy Hamiltonian $H_\lambda$.  Compared to
the canonical QCD theory, the confining interaction should
become more important if the scale $Q^2$ would
be smaller, and correspondingly the running coupling
constant $\overline{\alpha}(Q^2)$ becomes larger. This indicates
that there is an inverse correspondence between the effective
coupling constant $\alpha_\lambda$ in the low energy
Hamiltonian $H_\lambda$ and the running coupling constant
$\overline{\alpha}(Q^2)$ in the full QCD theory:
\begin{equation}	\label{invr}
	\alpha_\lambda \sim {1 \over \overline{\alpha}(Q^2)}~,
		~~~~ {\rm and}~~~~~  \lambda^2 \sim {1\over Q^2} .
\end{equation}
In other words, {\it the weak-coupling treatment of the low
energy confining Hamiltonian $H_\lambda$ may correspond to an
inverse strong-coupling expansion of the full QCD theory.
The similarity renormalization group approach provides an
implicative realization for such an expansion.} This may be the
inherent property why the nonperturbative QCD can be treated
as weak-coupling problem in the similarity renormalization
group scheme and why we can find the confining interaction
in this weak-coupling QCD formulation.

We also find from Table IV that the confining interaction plays
a more important role than the Coulomb interaction
in the determination of the quarkonium bound states.
This result is different from the usual understanding in the
nonrelativistic phenomenological description that the
dominant contribution in heavy quarkonium spectroscopy
is the Coulomb interaction.  This discrepancy can be
understood as follows.  The currently relativistic
light-front description for heavy quark system mostly
uses the heavy quark masses of $m_c=1.3 \sim 1.4$ GeV
and $m_b=4.8$ GeV or less (In Particle Data Group
\cite{PDB}, $m_b=1.0 \sim 1.6$ GeV and $m_c=4.1 \sim 4.5$
GeV). Thus, the heavy quarkonium binding energies,
$\overline{\Lambda}= M_H - 2m_Q$, might be positive
[the lowest charmonium ground state $M(\eta_c(1S))=2.98$
GeV, and the bottomonium $M(\Upsilon(1S)) = 9.46$ GeV].
Therefore, the Coulomb energy is obviously not important. The
dominant contribution for binding quarkonium states must come
from the nonperturbative balance between the kinetic energy
and the confining energy. While, in the nonrelativistic
phenomenological description, one used the larger
quark masses, $m_c > 1.8$ GeV and $m_b > 5.1$ GeV \cite{IGWS},
such that the binding energy is negative and therefore
the Coulomb interactions must be dominant in this picture.
Of course, on the light-front, the structure of the bound
state equation is different from the nonrelativistic
Schr\"{o}dinger equation.  There is no direct comparison.
A real solution to the above discrepancy may be obtained
after including the spin-splitting interactions
($1/m_Q$-corrections).

Now we can study the running behavior of the coupling
constant in the similarity renormalization group scheme.
Denote
\begin{equation}
	 \overline{\Lambda}=  \overline{\Lambda}
		(g_\lambda, \omega_\lambda, \lambda).
\end{equation}
The invariance of the binding energy $\overline{\Lambda}$ under
the similarity renormalization group transformation means that the
$\overline{\Lambda}$ determined from $H_\lambda$ and $H'_{\lambda'}$
must be the same for $\lambda \neq \lambda'$. Let $\lambda' = \lambda
+ \delta \lambda$, we obtain the corresponding similarity
renormalization group equation
\begin{equation}	\label{srge}
	\Big( \lambda {\partial \over \partial \lambda} + \beta
		{\partial \over \partial g_\lambda} + \gamma_\omega
		{\partial \over \partial \omega_\lambda} \Big)
	\overline{\Lambda} (g_\lambda, \omega_\lambda, \lambda)=0,
\end{equation}
where the quantity $\beta$ is the similarity renormalization group
$\beta$ function which is defined by
\begin{equation}	\label{beta1}
	\beta(g_\lambda) = \lambda {d g_\lambda
		\over d\lambda}\Bigg|_{\lambda
		=\lambda(g_\lambda)},
\end{equation}
and $\gamma_\omega$ is an anomalous dimension that describes the
running properties of the bound state wavefunction.
The $\beta$ function can be computed from eq.(\ref{running}),
\begin{eqnarray}
	\beta &=& - g_\lambda \Bigg(1 + {2b\over a + 2b \ln {\lambda
		\over \overline{\Lambda}}}\Bigg)\Bigg|_{\lambda
		=\lambda(g_\lambda)} \nonumber \\
	&\approx& - g_\lambda  ~~~~~( {\rm for~a~relatively~large}~~~
		\lambda >> \overline{\Lambda} ). \label{beta}
\end{eqnarray}

On the other hand, the running coupling constant
in full QCD theory is given by
\begin{equation}
	 t= \int_{g_s}^{\overline{g}} {d g' \over \beta(g')},
\end{equation}
where $	t= {1\over 2}\ln {Q^2 \over \mu^2}$, and $Q^2$ is a
space-like momentum (the same as $\lambda^2$). Since the
similarity renormalization group $\beta$ function of eq.(\ref{beta})
is determined in the physical sector of low energy QCD
dynamics, the low energy $\beta$ function of the running
coupling constant $\overline{g}(Q^2)$ in the full theory
should behave qualitatively the same.  With this assumption,
the $\beta(g)$ function in the above equation may take the same
form as eq.(\ref{beta}) in low momentum transfer (namely
$Q^2 << \Lambda^2_{QCD}$). This leads to
\begin{equation}
	\overline{g}^2 (Q^2) = g^2_s {\mu^2 \over Q^2 }~~,
	~~~~~Q^2 << \Lambda^2_{QCD},
\end{equation}
and $g^2_s = g^2_s(\mu^2)$ is a fixed coupling constant
at the hadronic mass scale $\mu^2$.  In terms of the running
coupling constant $\alpha=g^2 /4\pi$, we have
\begin{equation}	\label{sQ2}
	\overline{\alpha} (Q^2) = \alpha_s (\mu^2)
	{\mu^2 \over Q^2 } \equiv c_0 {\Lambda_{QCD}^2 \over Q^2},
\end{equation}
where $c_0= \alpha_s (\mu^2)\mu^2/\Lambda_{QCD}^2$. This
is consistent with eq.(\ref{invr}). Furthermore, we
see that the fixed point of the coupling constant under
renormalization group transformation is the origin of
$g$, and it is an infrared unstable (UV stable) fixed
point, in consistence with the asymptotic freedom of QCD.

To give a qualitative determination of the coefficient $c_0$,
we consider $\lambda =0.75 \sim 1.5$ GeV and $\overline{\Lambda}=0.2$
GeV, then $\lambda^2 = (14 \sim 56) \overline{\Lambda}^2 >>
\overline{\Lambda}^2$. Rewriting (\ref{running}) as the
same form of (\ref{sQ2}), we obtain:
\begin{equation}
	\alpha_\lambda = (1.0 \sim 1.5) ~{\overline{\Lambda}^2
		\over \lambda^2}.
\end{equation}
The corresponding $Q^2 \sim {1\over \lambda^2} <<
\overline{\Lambda}^2 \sim \Lambda_{QCD}^2$. From
(\ref{invr}), We may require that
\begin{equation}
	{\alpha_\lambda \over \overline{\alpha}(Q^2)}
		=  {Q^2 \over \lambda^2} .
\end{equation}
It follows that $c_0=1.0 \sim 1.5$, namely for $Q^2 <<
\Lambda_{QCD}^2$
\begin{equation}	\label{sQ21}
	\overline{\alpha}(Q^2) = (1.0 \sim 1.5)
		{\Lambda_{QCD}^2 \over Q^2}.
\end{equation}
This is just a qualitative estimation of the running
coupling constant in the full QCD theory in
low momentum transfer.  A more accurate result may be
obtained by exactly solving the $\beta$-function
of eq.(\ref{beta}). The running coupling constant
in high momentum transfer is given in the usual
perturbative QCD theory. A light-front perturbative
QCD calculation of the leading order running coupling
constant can be found from Ref.\cite{Hari93}. The result
is standard:
\begin{equation}	\label{lQ2}
	\overline{\alpha}(Q^2)={\alpha_s(\mu^2) \over 1+
		b_0 \alpha_s(\mu^2) \ln Q^2/\mu^2} = {12 \pi
		\over (33-2N_f) \ln Q^2/\Lambda_{QCD}^2} .
\end{equation}

Up to date, no one precisely knows how the QCD coupling
constant varies in low energy scale.  However,
it is interesting to see that the running coupling
constant given by eqs.(\ref{sQ2}) and (\ref{lQ2}) for the small
and large $Q^2$ respectively is indeed the basic assumption
of the Richardson $Q\overline{Q}$ potential \cite{Rich79}:
\begin{equation}	\label{QQp}
	V(Q^2) = - C_f {\overline{\alpha}(Q^2) \over Q^2} ,
\end{equation}
where
\begin{equation}
	\overline{\alpha}(Q^2) = {12 \pi \over (33-2N_f)
		\ln(1 + Q^2/\Lambda_{QCD}^2)} .
\end{equation}
The Richardson $Q\overline{Q}$ potential is proposed to
exhibit the asymptotic freedom of QCD in short distance and
a linear potential in large distance.  From eq.(\ref{QQp}),
we see that for large $Q^2 ~(Q^2 >> \Lambda_{QCD}^2)$,
\begin{equation}
	\overline{\alpha}(Q^2) \sim {12 \pi \over
		(33-2N_f) \ln Q^2/\Lambda_{QCD}^2},
\end{equation}
which reproduces the result of the asymptotic freedom of QCD.  The
corresponding potential is just the Coulomb potential. For small
$Q^2 ~(Q^2 << \Lambda_{QCD}^2) $,
\begin{equation}	\label{sQ22}
	\overline{\alpha}(Q^2)  \sim {12 \pi \over
		33-2N_f} ~ {\Lambda_{QCD}^2 \over Q^2},
\end{equation}
and the corresponding potential from (\ref{QQp}) becomes
\begin{equation}
	V(Q^2) \sim - {12 \pi C_f \Lambda_{QCD}^2 \over 33-2N_f}
		{1 \over Q^4}
\end{equation}
which is a Fourier transformation of the linear potential,
\begin{equation}
	V(r) = k r .
\end{equation}
The Fourier transformation of (\ref{QQp}) is the $Q\overline{Q}$
potential in coordinate space:
\begin{equation}
	V(r) = {8\pi \over 33-2N_f} \Lambda_{QCD} \Bigg(r\Lambda_{QCD}
		- {f(r\Lambda_{QCD}) \over r\Lambda_{QCD}} \Bigg) ,
\end{equation}
where $f(t) = \Big[1 - 4 \int_1^\infty {dq \over q}{e^{-qt} \over
[\ln(q^2-1)]^2 + \pi^2}\Big]$ \cite{Rich79}.  The Richardson potential
has successfully been used to describe quarkonium dynamics.

Comparing with eqs.(\ref{sQ2}) and (\ref{sQ22}),
we have from the Richardson $Q\overline{Q}$ potential
(with $N_f=3$ \cite{Rich79})
\begin{eqnarray}	\label{last}
	c_0 = {12 \pi \over (33-2N_f)}  =  1.4~.
\end{eqnarray}
This result agrees very well with eq.(\ref{sQ21}). Consequently,
although it is a very rough qualitative analysis,
the above result may imply that the confining Hamiltonian
derived from light-front HQET in the similarity
renormalization group scheme has also covered the dynamics
of linear confining potential.
As we have known, phenomenological potential quark models
based on the Richardson potential, the linear
plus Coulomb potential (also called the Cornell
potential) as well as the logarithmic potential
all give a good description of quarkonium dynamics \cite{Kwong}.
Hence, it should be not surprising if our QCD confining
Hamiltonian encompass the dynamic behavior of all these
potentials.

\section{Discussions on the full QCD via effective theory}

Thus far, the main ideas of the weak-coupling
treatment on nonperturbative QCD proposed in the recent
publication \cite{Wilson94} have, at least qualitatively,
been achieved for heavy quarkonium. The low energy
nonperturbative QCD theory is defined by the effective
low energy Hamiltonian $H_\lambda$.  The key to
solve this theory is to determine from the bound state
equation the $\lambda$-dependence of the effective
coupling constant $\alpha_\lambda$. Our result indicates
that the low energy effective QCD Hamiltonian exhibits an
alternative realization of the inverse strong coupling
expansion of the full QCD theory.  Thus, with a
suitable choice of the cutoff $\lambda$, the effective
coupling constant $\alpha_\lambda$ (as well as $g_\lambda$)
can be arbitrarily small. As an example, one may take
$\lambda$ to be a constituent gluon mass (about a half
of the glueball masses, such as the recent possible evidences
of $f_0(1500)$ and $\xi(2230)$ \cite{Glueball}),
\begin{equation}	\label{lam}
	\lambda :~ (0.75 \sim 1.5) ~{\rm GeV}.
\end{equation}
Then the effective coupling constant (with $\overline{\Lambda}
= 200$ MeV) is
\begin{equation}
	\alpha_e (\lambda^2) = 0.06 \sim 0.01
\end{equation}
which is very small. The residual interaction $H_{\lambda I}$
is expanded in terms of this small coupling constant so that the
corrections from $H_{\lambda I}$ can be perturbatively computed,
and the weak-coupling treatment of nonperturbative QCD is
explicitly realized.

Now the question is in what limit the effective theory can
return back to the full theory of QCD so we can ensure that
the present formulation is a complete consistent theory describing
low energy QCD dynamics. This question is also important in the
sense that with the cutoffs being introduced and the assumption of
gluon being massive, the gauge symmetry and rotational symmetry
may be broken down in the effective theory.  Then we must know
when all these symmetries can be restored. In ref.\cite{Wilson94},
we have argued that when $g_\lambda \rightarrow g_s$, the
effective theory must recover the full QCD dynamics so that
all symmetries are restored as well.  However, it is not clear
how this limit can be achieved.  Here we shall attempt to
answer this question from the heavy quarkonium solution.

We have used three cutoffs in this paper: the UV cutoff
$\Lambda$, the low energy cutoff $\lambda$ and the infrared
longitudinal momentum cutoff $\epsilon$. The UV cutoff
is renormalized away in ordinary perturbation theory
so that there is no any explicit $\Lambda$ dependence in our
formulation. The longitudinal infrared cutoff is automatically
cancelled in all the physical sectors, due mainly to the
gauge invariance as we have seen from the calculations
throughout this work. The final
formulation only contains the low energy cutoff $\lambda$.
This $\lambda$ dependence is essentially associated
with nonperturbative QCD dynamics.  However, the similarity
renormalization group invariance on physical observations
allows us to further remove away the $\lambda$ dependence
in physical sectors.  We may define that the naive $\lambda
\rightarrow 0$ limit brings the effective theory back to
the full theory of QCD.

How can the limit of $\lambda \rightarrow 0$ theoretically
bring the effective theory back to the full QCD theory?
Firstly, recall that introducing the low energy cutoff
in the effective theory is based on the assumption
that gluon is massive in the low energy domain.  Then
the limit of $\lambda \rightarrow 0$ is only allowed
when the gluon mass goes to zero.  With this limit
the broken gauge symmetry due to the massive gluons is
now restored. Secondly, once the cutoff $\lambda$ is
renormalized away, there is no any explicit cutoff
dependence in the theory. Therefore the broken
Lorentz symmetry due to the use of the explicit cutoff
must be restored as well. Furthermore,
the limit $\lambda \rightarrow 0$ corresponds to
$Q^2 \rightarrow \infty$, where all gluons and quarks
become current ones.  Once all symmetries are restored
and the current picture reemerges, the resulting theory
should be the full QCD theory.

We may first check what happens if we perform the
same procedure to QED for positronium. With $\lambda
\rightarrow 0$, the similarity renormalization approach
leads to an effective QED Hamiltonian in which the nonperturbative
part $H_{\lambda 0}$ only contains the Coulomb interaction,
and the remaining is the radiative correction $H_{\lambda I}$.
No confining interactions and no infrared divergences occur,
as expected. This is just the full QED theory used in the
description of QED bound states. We can explicitly examine
the above conclusion from eq.(\ref{sQQe}).  To do so, we may
first assume that the QED coupling constant is almost
$\lambda$-independent because it is always very weak
in the whole range of energy scale.  Thus, with $\lambda
\rightarrow 0$, we have from eq.(\ref{sQQe}),
\begin{equation}
	 {\cal E}_{nonc.} = 0,
	  ~~~  {\cal E}_{Coul.} = - {g^2\over \pi^2} C_f
		{\sqrt{\pi} \omega \overline{\Lambda}},
\end{equation}
namely, only Coulomb force contributes, and the confining
interaction disappears, while the infrared divergence
in the self-energy correction does not occur for $\lambda=0$.
Combining with the kinetic energy (where the term
$\overline{\Lambda}^3/m_Q$ in ${\cal E}_{kin}$ is very small
so that it can be neglected), the totally binding energy
is given by
\begin{equation}
	\overline{\Lambda} = {3\over 2m_Q} \omega^2 - {g^2\over
		2\pi^2} C_f {\sqrt{\pi}\omega} .
\end{equation}
Minimizing $\overline{\Lambda}$ with respect of $\omega$,
we obtain
\begin{equation}	\label{bohr1}
	\omega_0 = {g^2 m_Q \over 6 \pi^{3/2}} C_f~,
	~~~ \overline{\Lambda} = - {g^4 m_Q \over 24 \pi^3} C_f^2 .
\end{equation}
The above result can be rewritten as an exact solution of
QED for the positronium ground state in the
nonrelativistic limit.  If we let the color factor $C_f=1$
and $m_Q \rightarrow m_e$, $g \rightarrow e$, eq.(\ref{bohr1})
can be reexpressed in terms of the Bohr radius and Bohr energy
(in unit $h=1$):
\begin{equation}
	a_0 = {\sqrt{\pi} \over 2 \omega_0}
		= {3\pi^2 \over m_e e^2} ~,
	~~~ E_0 = \overline{\Lambda} = - {\alpha \over 2 a_0}~~,
	~~~~~(\alpha = {e^2 \over 4\pi})
\end{equation}
where the Bohr radius is redefined since we use a Gaussian-type
trial wavefunction which is not the same as the exact hydrogen
atomic ground state wavefunction.

In fact, if we did not assume the $\lambda$-independent of
the QED coupling constant, we would obtain, except for a color
factor, the same similarity renormalization group $\beta$
function of eq.(\ref{beta}).  Thus, the
fixed point of QED running coupling constant, $e=0$,
becomes infrared unstable, which is inconsistent with
the well-known perturbative QED result of being an
infrared stable fixed point. It implies that the similarity
renormalization scheme can be applied to QED only for
$\lambda=0$.  With $\lambda=0$, our formulation reproduces
the well-known method for solving QED bound states, namely, the
nonperturbative bound states is determined by solving
the Schr\"{o}dinger equation with the Coulomb potential
(because the effective Hamiltonian $H_{\lambda 0}$ only
contains the Coulomb interaction at $\lambda=0$) and the
remaining relativistic radiative corrections (described
here by $H_{\lambda I}$) can then be systematically
computed in perturbation theory. In other words, the
limit $\lambda \rightarrow 0$ brings the effective theory
back to the full theory in QED.

However, unlike the QED, we cannot assume that the coupling
constant $g$ in QCD is independent of the scale $\lambda$.
In other words, one cannot freely take the limit of $\lambda
\rightarrow 0$.  With  $\lambda$ being decreased, $g_\lambda$
becomes larger and larger.  Thus, the resulting $H_{\lambda 0}$
containing the Coulomb interaction alone is not sufficient
to describe QCD bound states since the corrections
from $H_{\lambda I}$ cannot be handled perturbatively.
Therefore we are in practice unable to write down the full QCD
theory in the weak-coupling formulation. The weak-coupling
treatment of QCD in the limit $\lambda \rightarrow 0$
is no longer valid.  Indeed, we find that with
a small $\lambda$ ($\lambda < 0.4$ GeV), eq.(\ref{sQQe})
has no solution. In other words, the so-called full QCD theory
with massless gluon in low energy domain may only be
formally interesting.  To reproduce the acceptable
hadronic properties with a small $\lambda$ value, we must
include more complicated higher order contributions
from $H_{\lambda I}$ into the nonperturbative bound state
equation. Since the lowest bound value of $\lambda$ is
the constituent gluon mass, a finite $\lambda$, namely
a nonzero constituent gluon mass, has effectively moved
the nonperturbative contribution in the higher order
processes into the low energy two-body confining
interaction.  The fact that the confining interaction
dominates the binding dynamics of the quarkonium bound
states at a finite $\lambda$ (about 1 GeV)
is indeed an evidence why nonperturbative QCD can
be treated as a weak-couple problem in the present formulation.
The limit $\lambda \rightarrow 0$ that can bring the
effective theory back to the full theory is only
implicative in QCD.

On the other hand, the result from quarkonium ground states
seems to tell us that with the larger $\lambda$, the smaller
the effective coupling constant can be reached. But this
does not imply that the weak-coupling treatment to
nonperturbative QCD works better for a larger $\lambda$.
The scale dependence of the wavefunction provides
a restriction on the value of $\lambda$.  For quarkonia,
$\omega_\lambda$ is decreased with increasing $\lambda$.
However, $\omega_\lambda$ is proportional to the mean value of
the (transverse) momentum square of the quark inside the
heavy quarkonia which characterizes the size of hadrons.
Therefore it should not be too large or too small in the best
description for bound states. For the range of eq.(\ref{lam}),
we have,
\begin{equation}
 	\omega_{\lambda 0} = 0.24 \sim 0.2 ~{\rm GeV}.	\label{omm}
\end{equation}
Correspondingly,
\begin{equation}
	\langle r \rangle \sim {1 \over \omega_{\lambda 0}}
		= 0.8 \sim 1.0 ~{\rm fm} ,
\end{equation}
which gives a resonable quarkonium size. Therefore, the
true low energy QCD theory is determined by $H_\lambda$ with
$\lambda$ being around the hadronic mass scale.

Although our formulation thus explicitly involves $\lambda$,
by the requirement of the similarity renormalization group
invariance, all the physical observables computed in this
effective theory can still be $\lambda$ independent. The naive
limit of $\lambda \rightarrow 0$ that brings the effective
theory back to the full QCD theory may imply that
the final physical results calculated from $H_\lambda$ could
also be gauge and Lorentz invariant, although $H_\lambda$
itself does not have these symmetries (consequently the hadronic
wavefunctions must also be $\lambda$-dependent).

In conclusion, the weak-coupling treatment approach
to hadronic bound states in QCD can work well with the
cutoff $\lambda$ being around the hadronic mass scale.
The well-defined bound state approach in QED is a special
case ($\lambda=0$) of the similarity renormalization
group approach.  The whole idea of the weak-coupling
treatment to nonperturbative QCD via the similarity
renormalization group approach is originally motivated from
the bound state description of QED \cite{Wilson94}.
Now, a consistent connection between QCD and QED and
their differences in low energy domain is explicitly
examined.

\section{Summary}

In this last section, we shall summarize the general
formulation and the main results we have obtained, and
then briefly discuss the further works.

To realize the weak-coupling treatment of nonperturbative
QCD recently proposed by Wilson and his
collaborators \cite{Wilson94},
we have studied explicitly the heavy quark
bound state problem, based on the recently developed light-front
heavy quark effective theory of QCD \cite{Zhang95,Cheung95}.
Firstly, we have used the similarity renormalization group approach
\cite{Wilson94,Glazek94} to derive the effective confining
Hamiltonian in the low energy scale for heavy quarks in heavy
mass limit. To make the similarity renormalization
approach practically manable, we have introduced a local
cutoff scheme (\ref{ctf}) to the bare QCD (and the effective
heavy quark) Hamiltonian, which simplifies the cutoff
scheme in \cite{Wilson94}.  Meanwhile we have also introduced
a simple smearing function $f_{\lambda ij}$ (\ref{sm2})
to the similarity renormalization group approach which
further simplifies the original formulation of
\cite{Wilson94}. The resulting low-energy effective QCD
Hamiltonian of heavy quark interactions exhibits the coexistence
of a confining interaction and a Coulomb interaction on the
light-front.

The realization of the weak-coupling treatment to nonperturbative
QCD dynamics is very much based on the reseparation of the
low energy effective Hamiltonian $H_\lambda$ into a
nonperturbative part $H_{\lambda 0}$ and the remaining as
a perturbative term $H_{\lambda I}$, and also on the use of
the constituent picture, as we have seen throughout
the present work. The use of the constituent picture
in light-front field theory allows us to expand the
heavy hadrons only with the
valence quark Fock space.  The light-front heavy quark
effective theory also largely simplifies the structure of
the heavy hadron bound state equations [see (\ref{QQbse})
and (\ref{Qqbse})].

A true realization of the weak-coupling approach to
nonperturbative QCD can be obtained after solving the light-front
bound state equations, from which one can determine the
$\lambda$-dependence of the effective coupling constant
in $H_\lambda$, as a solution of the similarity
renormalization group invariance.  We have used a well-behaved
light-front wavefunction ansatz (a Gaussian form) to
analytically solve the quarkonium bound state
equation and determine the scale-dependence
of the effective coupling constant. We have also shown
that the effective coupling constant at the hadronic
scale $\lambda$ can be arbitrarily small. Thus,
the possible weak-coupling treatment
to nonperturbative QCD proposed in ref.\cite{Wilson94} is
explicitly achieved.

The results obtained in this paper is very optimistic.
First, the $\lambda$-dependence of the effective
coupling constant determines the similarity renormalization
group $\beta$ function, from which some qualitative running
behavior of the coupling constant in low energy domain
is obtained. The running coupling constant (\ref{sQ21})
is qualitatively the same one used in the
successful Richardson $Q\overline{Q}$ potential for small
momentum transfer (large distance). The later is a basic
assumption to ensure the existence of a linear confining
$Q\overline{Q}$ potential in large distance, which
is now obtained from QCD in the present work. A light-front
picture of quark confinement from QCD is naturally manifested
in $H_{\lambda 0}$. It encompasses the general properties of these
phenomenological confining potentials, the Richardson potential,
the linear plus Coulomb potential and the logarithmic
potentials used in the phenomenological description
of quarkonium dynamics.

The weak-coupling treatment can be realized for nonperturbative
QCD because the similarity renormalization group approach
with a finite $\lambda$ has extracted the confining interaction
from the higher order nontrivial quark-gluon interactions
into $H_{\lambda 0}$. Equivalently speaking, the similarity
renormalization group approach implicatively provides an inverse
strong interaction expansion of QCD via the low energy cutoff
scale $\lambda$. As a physical consequence, the confining
interaction plays a dominant role in hadronic bound states,
as we have seen in Section VII.
The weak-coupling treatment of nonperturbative
QCD is manable for $\lambda$ being around the hadronic
mass scale.  The similarity renormalization group invariance
can remove the $\lambda$-dependence in all the physical
observables obtained from the effective $H_\lambda$.
The possible connection of the effective theory to the
full QCD theory in the limit $\lambda \rightarrow 0$ may further
imply that the final physical results obtained from
$H_\lambda$ could be gauge and Lorentz invariant. Meanwhile, we
have also shown that in consistence with the behavior of the
QED $\beta$ function, the similarity renormalization
approach can be applied to QED only if $\lambda=0$.
Therefore, the confining picture obtained for QCD does
not exist in QED but the well-known QED bound state
method is reproduced. Thus, the consistency of the
weak-coupling formulation has been qualitatively examined.

The applications of the present theory to heavy hadron
spectroscopy and various heavy hadron decay processes
can be simply achieved by numerically solving the bound
state equations (\ref{QQbse}) and (\ref{Qqbse}), and
by further including the $1/m_Q$ corrections (which naturally
leads to the spin splitting interactions).  These will be
presented in a forthcoming paper \cite{Zhang96}. Finally,
we should analyze the systematical approximations
used in the whole computations in this paper,
and then discuss the further works along this direction.

The entire derivations presented in this paper are purely
based on the first principle of QCD. Applying to heavy hadrons,
we took the heavy mass limit so that the QCD is reduced to
HQET but the first order kinetic energy has been also included
in the leading order Hamiltonian for heavy quarkonia.
In the forthcoming paper \cite{Zhang96}, the $1/m_Q$
corrections (which contains all the spin splitting
interactions) will be considered in the effective
Hamiltonian $H_{\lambda I}$ in order to
compute the heavy quarkonium spectroscopy and the next
order correction to the bound states. These corrections
should not affect on the main conclusions obtained
in this paper since the small factor $1/m_Q$ plus the weak
coupling constant guarantee that these $1/m_Q$ corrections
can be computed perturbatively with respect to the
nonperturbative heavy hadron bound states.

The low energy QCD Hamiltonian $H_\lambda$ is
obtained by a similarity renormalization transformation
to the bare QCD Hamiltonian (bare HQET Hamiltonian for
heavy quarks).  With the idea of the weak-coupling
treatment to low energy QCD dynamics, the nonperturbative
part $H_{\lambda 0}$ is computed upon to the second order
in $H_\lambda$ and the bound states are truncated to
only including the valence quark Fock space. The higher order
corrections in $H_\lambda$ (included both the $1/m_Q$ corrections
and the radiative corrections) can be examined in the usual
Hamiltonian perturbation theory \cite{Zhang96}, which will also
provide a consistent check to the validity of the present
weak-coupling treatment.  With these corrections
computed in the Hamiltonian formulation, the contributions
from the higher Fock space will be naturally included.

Finally, instead of numerically solving the light-front
bound state equations, we used here a well-behaved
light-front wavefunction ansatz to determine the
bound state equation for heavy quarkonia.  A numerical
calculation of the light-front bound state equations
(\ref{QQbse}) and (\ref{Qqbse}) for the heavy hadronic
spectroscopy will be also presented in \cite{Zhang96}.
Such a numerical computation of the bound states
is much simpler in comparison to the lattice
QCD simulation \cite{Lepage91}, and it will also
directly give the hadronic wavefunctions in
physical space. The resulting wavefunctions of the
quarkonium bound states combining with the
systematic computations of the subsequent radiative and
$1/m_Q$ corrections thus provide a truly unified first-
principles QCD description for various heavy quarkonium
annihilation and production processes.

All the derivations presented in this paper are rigorous
QCD derivation in low energy domain. The extension of
the computations to heavy-light quark systems is
straightforward but is more attractive in the current
investigations on heavy hadrons. The extension of the
present work to light-light hadrons requires
a understanding of the chiral symmetry breaking mechanism
in QCD which is a new challenge to nonperturbative QCD
on the light-front.  Nevertheless, the present work has
provided a preliminary realization to the weak-coupling
treatment of nonperturbative QCD proposed recently by
Wilson et al. \cite{Wilson94}. The new research
along this direction is now in progress.

\acknowledgements

The author would like to thank H. Y. Cheng, C. Y. Cheung,
S. B. Lee, S. C. Lee, T. M. Yan, W. B. Yang and H. L. Yu
for many fruitful discussions, and specially H. Y. Cheng
for his carefully reading the manuscript. This work is supported
in part by National Science Council under Grant Nos. NSC
84-2816-M-001-012L.

\newpage


\newpage


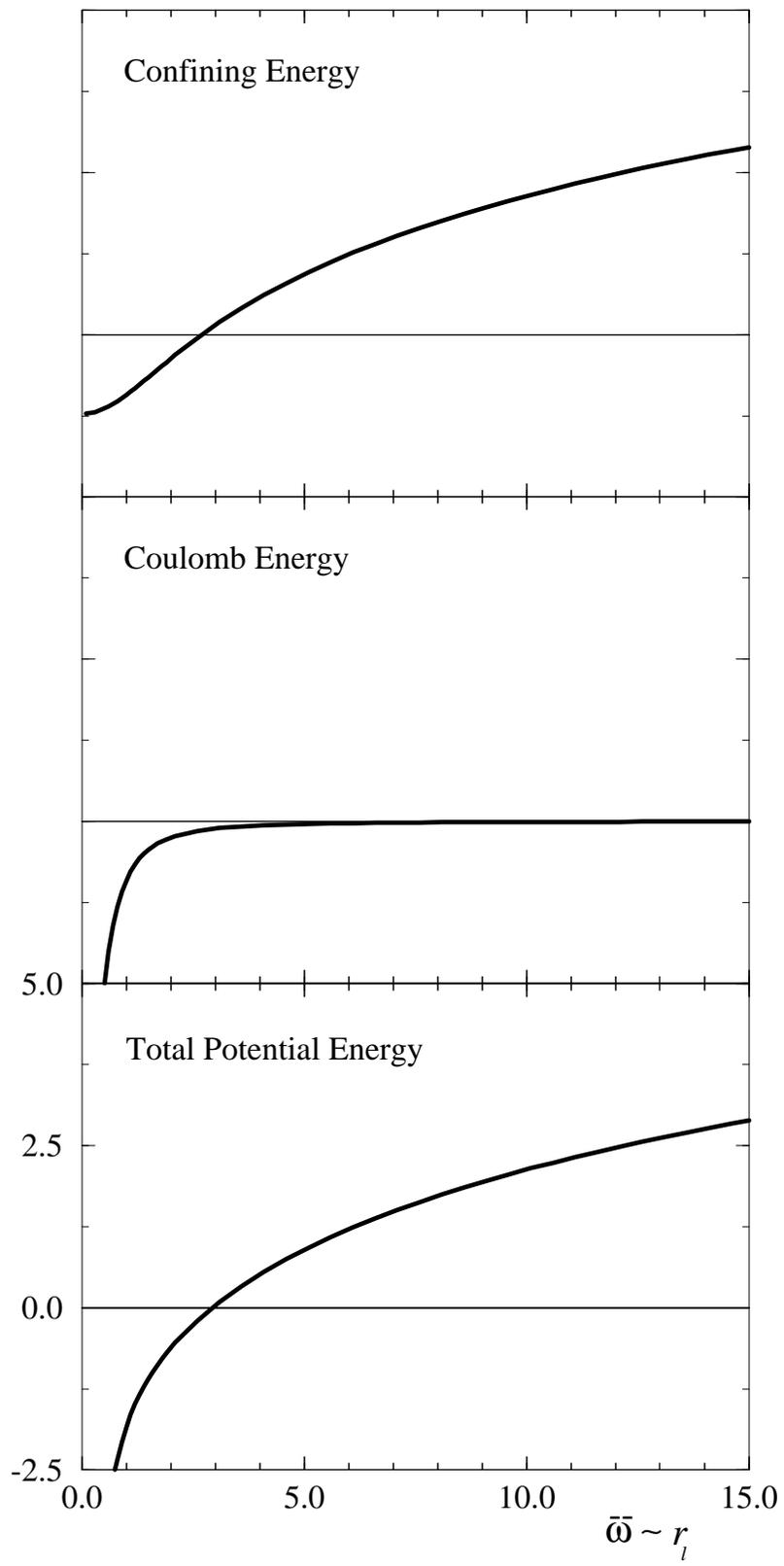
\begin{figure}
\caption[]{A plot of the confining potential energy,
the Coulomb potential energy and the total potential
energy as functions of the dimensionless variable
$\varpi$ but $\varpi$ is proportional to $\sim r_l$
via $\omega_\lambda$. The energies are scaled by the
factor ${g^2_\lambda \lambda^2 \over 2 \pi^2} C_f$.}
\end{figure}

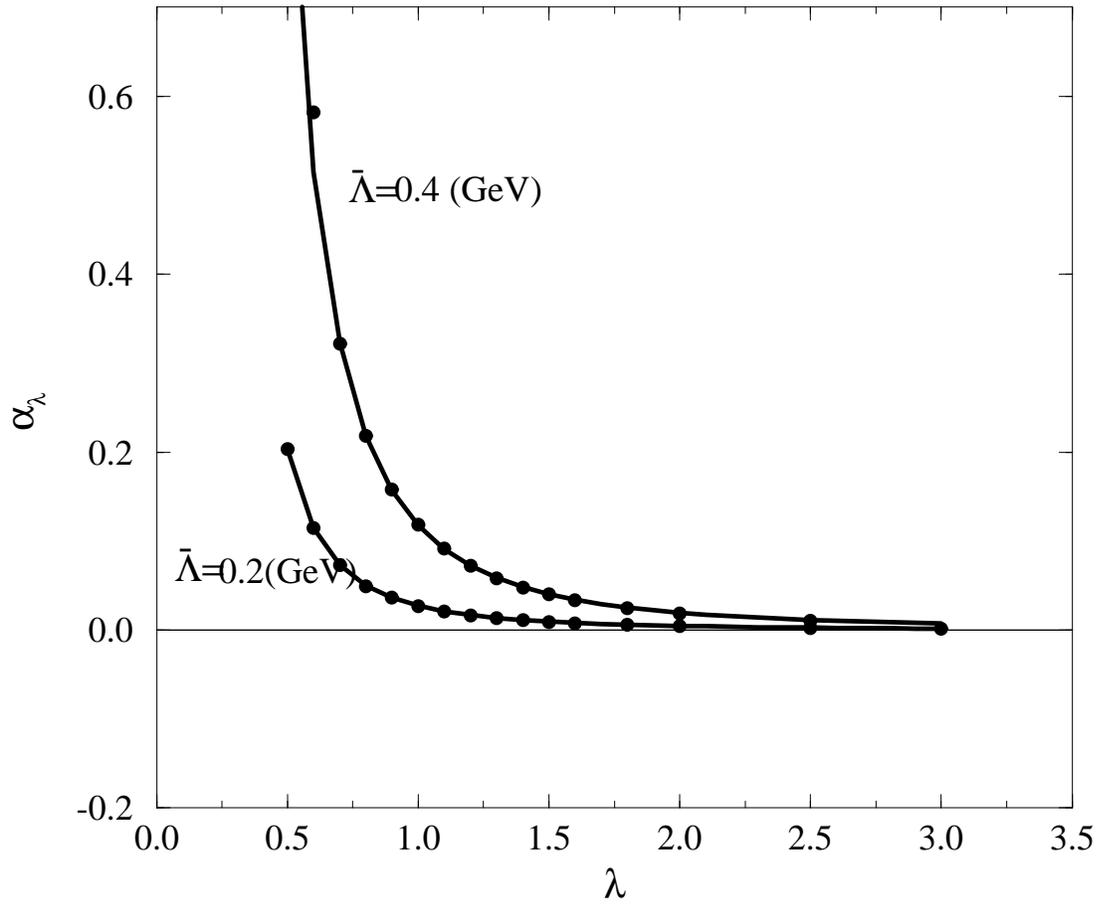
\begin{figure}
\caption[]{The $\lambda$-dependence of the effective
coupling constant $\alpha_\lambda$. The solid line
is given by the analytical result (\ref{running}),
and the dots are obtained by numerically
minimizing the quarkonium binding energy
(\ref{sQQe}). Here $\lambda$ is given in units of GeV.}
\end{figure}

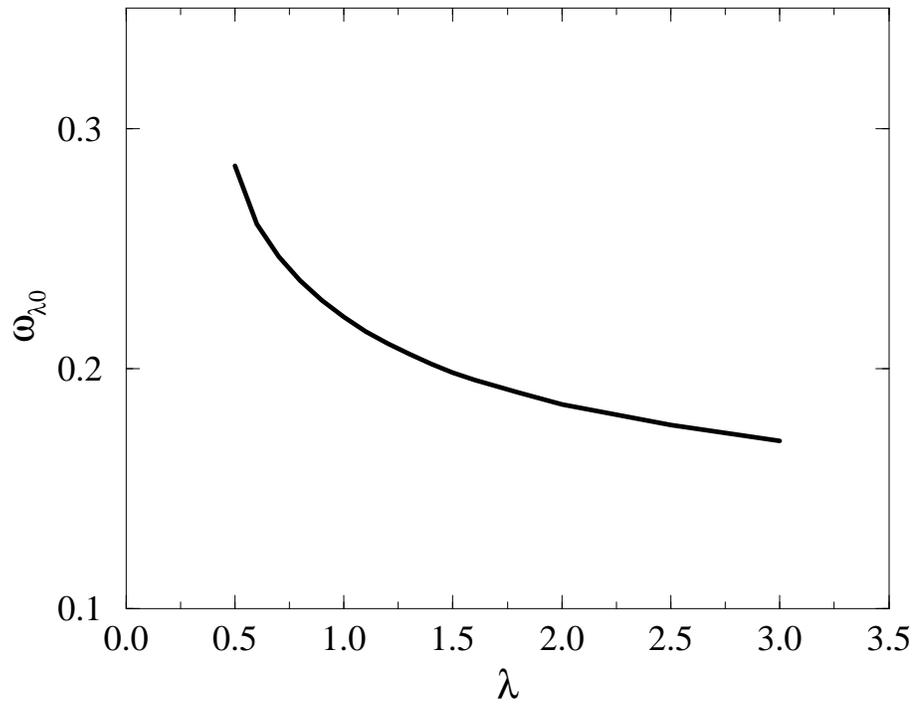
\begin{figure}
\caption[]{The $\lambda$-dependence of the wavefunction
parameter $\omega_{\lambda 0}$ which is the solution of
minimizing the quarkonium binding energy (\ref{sQQe}),
where $\lambda$ is given in units of GeV.}
\end{figure}

\newpage

\begin{center} TABLES

\vspace{0.5in}

\begin{tabular}{|c|c|c|c|c|c|}  \multicolumn{6}{c}{Table I.
Solution for charmonium ground state with $m_c=1.4$ GeV} \\
\hline\hline $\overline{\Lambda}$
(GeV) &~~~~ $\alpha_\lambda$~~~~& $\omega_{\lambda 0}$
(GeV) & \multicolumn{3}{c|} {${\cal E}_{kin}$(GeV$^2$) +
${\cal E}_{pot}$(GeV$^2$) = $2 \overline{\Lambda}^2$(GeV$^2$)} \\ \hline
0.2 & 0.02665 & 0.222 & ~~0.026836 ~~& ~~0.053173 ~~&~ 0.080009~ \\ \hline
0.3 & 0.06480 & 0.275 & 0.067902 & 0.112018 & 0.179920 \\ \hline
0.4 & 0.11831 & 0.314 & 0.130225 & 0.189781 & 0.320006 \\ \hline
\hline \end{tabular}

\vspace{0.5in}

\begin{tabular}{|c|c|c|c|c|c|}  \multicolumn{6}{c}{Table II.
Solution for bottomonium ground state with $m_b=4.8$ GeV} \\ \hline\hline
$\overline{\Lambda}$ (GeV) & $\alpha_\lambda$ & $\omega_{\lambda 0}$
(GeV) & \multicolumn{3}{c|} {${\cal E}_{kin}$(GeV$^2$) +
${\cal E}_{pot}$(GeV$^2$) = $2 \overline{\Lambda}^2$(GeV$^2$)} \\ \hline
0.15 & 0.029965 & 0.492 & ~~0.023397~~ &~~ 0.021602~~ &~ 0.044999~ \\
\hline 0.20 & 0.06795 & 0.623 & 0.050183 & 0.029816 & 0.079999 \\ \hline
0.25 & 0.1385 & 0.779 & 0.098074 & 0.026930 & 0.125004 \\ \hline
\hline \end{tabular}

\vspace{0.5in}

\begin{tabular}{|c|c|c|c|c|} \multicolumn{5}{c}{Table III. Some
numerical solution on the $\lambda$-dependence} \\
\multicolumn{5}{c}{of the weak coupling constant $\alpha_\lambda$.}\\
\hline\hline ~~& \multicolumn{2}{c|}{charmonium} &
\multicolumn{2}{c|}{bottomonium}\\ \hline $\lambda$ (GeV) &~~~~~
$\alpha_\lambda$~~~~~& ~$\omega_{\lambda 0}$ (GeV)~&~~~~~
$\alpha_\lambda$~~~~~& $\omega_{\lambda 0}$ (GeV) \\ \hline
0.75 & 0.05960 & 0.241 & ~~~~ & ~~ ~~\\ \hline
1.0 & 0.02665 & 0.222 & 0.06795 & 0.623 \\ \hline
1.5 & 0.00912 & 0.199 & 0.01607 & 0.478 \\ \hline
2.0 & 0.00441 & 0.185 & 0.00695 & 0.427 \\ \hline
\hline \end{tabular}

\vspace{0.5in}

\begin{tabular}{|c|c|c|c|c|c|c|c|c|} \multicolumn{9}{c}{Table IV. The
$\lambda$-dependence of various interactions to the binding energy}\\
\hline\hline $\lambda$ (GeV) & ~~~0.5~~~ & ~~~0.75~~~ & ~~~1.0~~~ & ~~~1.2
{}~~~ & ~~~1.4~~~ & ~~~1.8~~~ & ~~~2.0~~~& ~~~3.0~~~ \\ \hline
${\cal E}_{kin}$ (GeV$^2$) & 0.04 & 0.031& 0.027& 0.025& 0.023&
0.021& 0.02& 0.018\\ \hline
${\cal E}_{conf}$ (GeV$^2$) & 0.049 & 0.050& 0.053& 0.055&
0.057& 0.059& 0.06 &0.062 \\ \hline
${\cal E}_{Col}$ (GeV$^2$)  & -0.009 & -0.001& -0.001& 0.0&
0.0& 0.0 &0.0 & 0.0 \\ \hline \hline \end{tabular} \\ \ \\
\end{center}

\end{document}